\documentclass[a4paper,11pt]{article}
\usepackage{listings}
\usepackage{appendix}
\usepackage{pdfpages}
\usepackage{graphicx}
\usepackage{amsmath, amssymb}
\usepackage{url}
\usepackage{amsmath}
\usepackage{fancyvrb}
\usepackage{textcomp}
\usepackage{authblk}
\usepackage{verbatim}
\usepackage{fancyvrb}
\begin{document}

%
\title{On the Formal Semantics of the Cognitive Middleware AWDRAT\thanks{This is the draft version of the subject technical report.}}


\author[$\dag$]{Muhammad Taimoor Khan}
\author[$\dag$]{Dimitrios Serpanos}
\author[*]{Howard Shrobe}
\affil[$\dag$]{\{mtkhan, dserpanos\}@qf.org.qa}
\affil[*]{hes@csail.mit.edu}
\affil[$\dag$]{QCRI, Qatar}
\affil[*]{CSAIL, MIT, USA}

\renewcommand\Authands{ and }

\maketitle

\begin{abstract}
The purpose of this work is two fold: on one hand we want to formalize the behavior of critical components of the self generating and adapting cognitive middleware AWDRAT such that the formalism not only helps to understand the semantics and technical details of the middleware but also opens an opportunity to extend the middleware to support other complex application domains of cybersecurity; on the other hand, the formalism serves as a pre-requisite for our proof of the behavioral correctness of the critical components to ensure the safety of the middleware itself. However, here we focus only on the core and critical component of the middleware, i.e. Execution Monitor which is a part of the module ``Architectural Differencer'' of AWDRAT. The role of the execution monitor is to identify inconsistencies between runtime \emph{observations} of the target system and \emph{predictions} of the System Architectural Model. Therefore, to achieve this goal, we first define the formal (denotational) semantics of the \emph{observations} (runtime events) and \emph{predictions} (executable specifications as of System Architectural Model); then based on the aforementioned formal semantices, we formalize the behavior of the ``Execution Monitor'' of the middleware. 

\end{abstract}
\newpage
\tableofcontents
\newpage
\section{Calculus of AWDRAT}
Defending systems against cyber attack requires us to be able to rapidly and accurately detect that an attack has occurred.  Today's detection systems are woefully inadequate suffering from both high false positive and false negative rates.  There are two key reasons for this: First, the systems do not understand the complete behavior of the system they are protecting.  The second is that they do not understand what an attacker is trying to achieve.  Most such systems, in fact, are retrospective, that is they understand some surface signatures of previous attacks and attempt to recognize the same signature in current traffic.  Furthermore, they are passive in character, they sit back and wait for something similar to what has already happened to recur. Attackers, of course, respond by varying their attacks so as to avoid detection.

AWDRAT~\cite{Shrobe:2006} is a representative of a new class of protection systems that employ a different, active form of perception, one that is informed both by knowledge of what the protected application is trying to do and by knowledge of how attackers think.  It employs both bottom-up reasoning (going from sensors data to conclusions about what attacks might be in progress) as well as top-down reasoning (given a set of hypotheses about what attacks might be in progress, in focuses its attention to those events most likely to significantly help in discerning the ground truth).

There are two dimensions along which detection systems can be characterized.  The first is the distinction between profile and model based approaches.  The other dimension is the distinction between looking for matches to bad behavior or deviations from good.  This gives four quadrants, each with unique strengths and weaknesses.  For example, the bulk of our sensors are model-based and look for matches to bad behavior; signature based systems are in this category.  The advantage is that when a match occurs, you know what has happened; i.e. these systems have high diagnostic resolution.  But they also lack robustness; if they don't have a model of an attack, and there are always novel attacks, then they will fail to detect it.  On the other hand, there are a class of detectors that use employ machine learning techniques on labeled training data to build statistical profiles of attacks.  These systems tend to be a bit more robust than model based systems, since the machine learning techniques tend to generalize from the data presented.  However, they make up for this by a loss of diagnostic resolution.
The third quadrant involves building a statistical profile of normal behavior, detecting deviations from the profile.  Such anomaly detectors are yet more robust, since they don't depend on prior knowledge of the form of the attack, but they afford even less diagnostic resolution.  When things go wrong, all you know is that something out of the ordinary has happened; whether that something is malicious or not isn't known.

AWDRAT sits in the fourth quadrant: It has a model of normal behavior; when the application deviates from the behavior prescribed by that model, it employs diagnostic reasoning techniques~\cite{Shrobe:1979} to further isolate and characterize the failure.  It has both greater robustness and higher diagnostic resolution.  But it achieves this only through the construction of a far more complex model.

AWDRAT has an active model of normal behavior, namely an executable specification (aka System Architectural Model) of the application~\cite{Shrobe:2006}.  This executable specification consists of a decomposition into sub-modules and pre- and post-conditions for each sub-module.  In addition, data-flow and control-flow links connect the sub-modules, specifying the expected flow of values and of control. The pre- and post-conditions are arbitrary first-order statements about the set of data values that flow into and out of the sub-modules.

AWDRAT runs this executable specification in parallel with the actual application code, comparing their results at the granularity and abstraction level of the executable specification. (This is therefore a special case of the standard fault tolerance technique of running multiple versions of the same code and comparing their results.)   The executable specification is hierarchical, allowing flexibility in the granularity of the monitoring.  When threats are not expected, the executable specification is run at a high level of abstraction, incurring less overhead, but requiring more diagnostic reasoning should the program diverge from the prescribed behavior of the executable specification.  In times of heightened threat, the executable specification can be elaborated to a greater degree, incurring more overhead, but providing more containment.

Optionally, the model can also include models for suspected incorrect behaviors of a component, allowing the diagnostic reasoning to characterize the way in which a component might have misbehaved.  A diagnosis is then a selection of behavioral modes for each component of the specification such that the specification predicts the observed misbehavior of the system.

The rest of the paper is organized as follows: in Section~\ref{sec:sam} we discuss syntax of the System Architectural Model followed by the formalization of semantics of the critical syntactic domains model in Section~\ref{sec:spec-semantics}. Section~\ref{sec:monitor-semantics} formalizes the semantics of the execution monitor. Finally, we conclude in Section~\ref{sec:conclusions}. Appendices~\ref{sec:sam-syntax} and ~\ref{sec:example} give the formal syntactic grammar and an example System Architectural Model respectively.

\section{Syntax of System Architectural Model}\label{sec:sam}
An AWDRAT model is built from several related following forms which represent corresponding high-level syntactical domains of the model. Note, we only discuss selected domains here, for complete syntactic domains and their elements, please see Appendix~\ref{sec:sam-syntax}.

\begin{enumerate}
\item A description of a component type consists of 
\begin{enumerate}
\item its interface
\begin{itemize}
\item a list of inputs
\item a list of its outputs
\item a list of the resources it uses (e.g. files it reads, the code in memory that represents this component, etc)
\item list of subcomponents required for the execution of the subject component
\item a list of events that represent entry into the component (usually just one)
\item a list of events that represent exit from the component (usually just one)
\item a list of events that are allowed to occur during any execution of this component
\item a set of conditional probabilities between the possible modes of the resources and the possible modes of the whole component
\item a list of known vulnerabilities occurred to the component
\end{itemize}
\item and a structural model which is a list of sub-components some of which might be splits or joins of
\begin{itemize}
\item data-flows between linking ports of the sub-components (outputs of one to inputs of another)
\item control-flow links between cases of a branch and a component that will be enabled if that branch is taken	
\end{itemize}
\end{enumerate}
The description of the component type is represented by syntactical domain ``StrMod'' which is defined as follows:
\begin{tabbing}
StrMod ::= \textbf{define-ensemble} \=CompName
\\\> \textbf{:entry-events} \textbf{:auto} $|$ (EvntSeq)
\\\> \textbf{:exit-events} (EvntSeq)
\\\> \textbf{:allowable-events} (EvntSeq)
\\\> \textbf{:inputs} (ObjNameSeq)
\\\> \textbf{:outputs} (ObjNameSeq)
\\\> \textbf{:components} (CompSeq)
\\\> \textbf{:controlflows} (CtrlFlowSeq)
\\\> \textbf{:splits} (SpltCFSeq)
\\\> \textbf{:joins} (JoinCFSeq)
\\\> \textbf{:dataflows} (DataFlowSeq)
\\\> \textbf{:resources} (ResSeq)
\\\> \textbf{:resource-mapping} (ResMapSeq)
\\\> \textbf{:model-mappings} (ModMapSeq)
\\\> \textbf{:vulnerabilities} (VulnrabltySeq)
\end{tabbing}

\begin{description}
\item[Example 1:] The specification of the component \verb|maf-editor| is given below. In detail, the specification says that the component
\begin{itemize}
\item is top level component and hence starts automatically and thus requires no \verb|entry-event|,
\item requires no \verb|inputs|
\item results in \verb|the-model| as an output
\item has four subcomponents, i.e. \verb|startup|, \verb|create-model|, \verb|create-events| and \verb|save| which have corresponding types and also have both \verb|normal| and \verb|compromised| behaviors
\item has control and data flows as described
\item has an access to two resources, i.e. \verb|imagery| and \verb|code-files| which have corresponding probabilities of being in a \verb|normal| and \verb|hacked| mode
\item has model mappings of the above resources to the subcomponents as described in \verb|model-mappings| and
\item has two vulnerabilities, i.e. \verb|reads-complex-imagery| and \verb|loads-code| for the resources \verb|imagery| and \verb|code-files| respectively.
\end{itemize}

\begin{verbatim}
(define-ensemble maf-editor
    :entry-events :auto
    :inputs ()
    :outputs (the-model)
    :components 
    ((startup :type maf-startup :models (normal compromised))
    	(create-model :type maf-create-model :models (normal compromised))
    	(create-events :type maf-create-events :models (normal compromised))
    	(save :type maf-save :models (normal compromised)))
    				
    :controlflows ((before maf-editor before startup)
    				(after startup before create-model))
    				
    :dataflows ((the-model create-model the-model create-events)
    			(the-model create-events the-model save)
    			(the-model save the-model maf-save-model))
    			
    :resources ((imagery image-file (normal .7) (hacked .3)) 
    			(code-files loadable-files (normal .8) (hacked .2)))
    			
    :resource-mappings ((startup imagery)
    					(create-model code-files)
    					(create-events code-files)
    					(save-model code-files))
    					

    :model-mappings ((startup normal ((imagery normal)) .99)
		     (startup compromised ((imagery normal)) .01)
		     (startup normal ((imagery hacked)) .9)
		     (startup compromised ((imagery hacked)) .1)
		     
		     (create-model normal ((code-files normal)) .99)
		     (create-model compromised ((code-files normal)) .01)
		     (create-model normal ((code-files hacked)) .9)
		     (create-model compromised ((code-files hacked)) .1)
		     
		     (create-events normal ((code-files normal)) .99)
		     (create-events compromised ((code-files normal)) .01)
		     (create-events  normal ((code-files hacked)) .9)
		     (create-events compromised ((code-files hacked)) .1)
		     
		     (save normal ((code-files normal)) .99)
		     (save compromised ((code-files normal)) .001)
		     (save normal ((code-files hacked)) .01)
		     (save compromised ((code-files hacked)) .999))
    
    :vulnerabilities ((imagery reads-complex-imagery)
		      (code-files loads-code)
		      ))
\end{verbatim}
\end{description}
\item Behavioral specification of a component (a component type may have one normal behavioral specification and many abnormal behavioral specifications, each one representing some failure mode) which has
\begin{itemize}
\item  inputs and outputs
\item  preconditions on the inputs (logical expressions involving one or more of the inputs)
\item  postconditions (logical expressions involving one or more of the outputs and the inputs)
\item  allowable events during the execution in this mode
\end{itemize}
The behavioral specification of a component is represented by a corresponding syntactical domain ``BehMod'' as follows:

\begin{tabbing}
BehMod ::= \textbf{defbehavior-model} \=(CompName \textbf{normal} $|$ \textbf{compromised})
\\\> \textbf{:inputs} (ObjNameSeq)
\\\> \textbf{:outputs} (ObjNameSeq)
\\\> \textbf{:allowable-events} (EvntSeq)
\\\> \textbf{:prerequisites} (BehCondSeq)
\\\> \textbf{:post-conditions} (BehCondSeq)
\end{tabbing}

\begin{description}
\item[Example 2:] In the following first we give the structure of a component \verb|maf-create-model| (which is one of the submodule as stated in the previous specification example) and then give the behavioral specification of the component. The structure of the component is defined as follows:

\begin{verbatim}
(define-ensemble maf-create-model
    :entry-events (create-mission-action-action-performed)
    :exit-events (mission-builder-submit)
    :allowable-events (create-mission-builder-with-client-panel
		       create-mission-builder
		       create-mission-builder-with-hash-table
		       mission-builder-submit
		       (set-initial-info exit (the-model nil))
		       create-mission-action-action-performed
		       retrieve-info
		       create-mission-action-action-performed
		       (set-initial-info entry)
		       )
    :inputs ()
    :outputs (the-model))
    
\end{verbatim}
In the following we define the legal and illegal (compromised) behaviors of the component. For example, the specification of a legal (normal) behavior of the component says that as a normal behavior the component
\begin{itemize}
\item requires no input as specified by the clause \verb|inputs|
\item has \verb|the-model| output and also
\item no \verb|prerequisite| of the component but
\item guarantees that the object \verb|mission-builder| of \verb|the-model| are consistent.
\end{itemize}
The corresponding normal behavior is defined as:
\begin{verbatim}
(defbehavior-model (maf-create-model normal)
    :inputs ()
    :outputs (the-model)
    :prerequisites ()
    :post-conditions ([dscs ?the-model mission-builder good])
    )

(defbehavior-model (maf-create-model compromised)
    :inputs ()
    :outputs (the-model)
    :prerequisites ()
    :post-conditions ([not [dscs ?the-model mission-builder good]])
    )

\end{verbatim}
Similarly, the compromised behavior of the component is also described above. For further details on the behavioral specification of the other components, please see Appendix~\ref{sec:sam-syntax}.
\end{description}

\item Model of a resource type contains
\begin{itemize}
\item possible modes
\item prior probabilities of being in each mode
\item attack types to which it is vulnerable
\end{itemize}
The syntactical domain ``ResModMap'' represents the model of a resource type
\begin{tabbing}
ResModMap ::= \=ResName \textbf{normal} $|$ \textbf{hacked} FVal 
\\\>$|$ ((ResName \textbf{normal} $|$ \textbf{hacked})) FVal\\
\end{tabbing}
where ``FVal'' represents the float values for probabilities.

The trust model of the resources is specified in example 1 above by the clauses \verb|:resources| and \verb|:resource-mappings|.
\item Attack Model
\begin{itemize}
\item a list of types of attacks that are being anticipated and the prior probability of each
\item a list describing how each attack type can effect that mode of a resource
\item a set of logical rules expressing the conditional probabilities between attack types and resource modes
\end{itemize}

The attack models are presented by the syntactic domain ``AtkMod'' while the corresponding attack rules are specified by the syntactic domain ``AtkRule'' as given below respectively:

\begin{tabbing}
AtkMod ::= \textbf{define-}\=\textbf{attack-model} AtkModName
\\\> \textbf{:attack-types} (AtkTypeSeq)
\\\> \textbf{:vulnerability-mapping} (AtkVulnrabltyMapSeq)
\end{tabbing}

\begin{tabbing}
AtkRule ::= \textbf{defrule} \=AtkRulName (\textbf{:forward})
\\\> \textbf{if} AtkCondSeq
\\\> \textbf{then} AtkConsSeq
\end{tabbing}

\begin{description}
\item[Example 3:] The example attack model \verb|maf-attacks| specifies the two attacks
\verb|hacked-image-file-attack| and \verb|hacked-code-file-attack| with some
probabilities as specified in the following.

\begin{verbatim}
(define-attack-model maf-attacks
    :attack-types (	(hacked-image-file-attack .3) 
    				(hacked-code-file-attack .5))
    :vulnerability-mapping 
    			((reads-complex-imagery hacked-image-file-attack)
    					(loads-code hacked-code-file-attack)))
			    
\end{verbatim}
Furthermore,  the two attacks are mapped to the corresponding vulnerabilities
\verb|reads-complex-imagery| and \verb|loads-code| respectively.

Additionally, the corresponding one attack rule \verb|bad-image-file-takeover| says that \verb|if| we have the contextual resource \verb|?ensemble| and \verb|type-of-resource| is \verb|image-file| and the \verb|resource-might-have-been-attacked| with \verb|hacked-image-file-attack| \verb|then| it is highly probable (\verb|.9|) that the resource
has been \verb|hacked| by \verb|hacked-image-file-attack| as given below:
\begin{verbatim}

(defrule bad-image-file-takeover (:forward)
  if [and [resource ?ensemble ?resource-name ?resource]
	  [resource-type-of ?resource image-file]
	  [resource-might-have-been-attacked ?resource 
	  		hacked-image-file-attack]]
	  
  then [and [attack-implies-compromised-mode 
  				hacked-image-file-attack ?resource hacked .9 ]
	    [attack-implies-compromised-mode 
	    			hacked-image-file-attack ?resource normal .1 ]])
	    
\end{verbatim}

 
 \end{description}

\end{enumerate}

Further details on the example of the model, please see Appendix~\ref{sec:example}. However, 
the corresponding syntactic details of the elements of the above syntactic domains are explained in the corresponding subsections of the next section. However, for the general syntax of the domain, please see Appendix~\ref{sec:sam-syntax}.

\section{Semantics of System Architectural Model}\label{sec:spec-semantics}
In this section, we first give the definition of semantic algebras and then discuss informal description and the formal semantics of the core constructs of the System Architectural Model.

\subsection{Semantic Algebras}
The definition of a formal denotational semantics is based on a collection of data structures. 
\emph{Semantic domains} represent set of elements that share some common properties. A semantic domain is accompanied by a set of operations as functions over the domain. A domain and its operations together form a \emph{semantic algebra}~\cite{Schmidt86}. In the following we enlist the semantic domains and their corresponding operations. Some operations are defined and some are just declared for the purpose of completeness of this document.

\subsubsection{Truth Values}
The truth values are represented by the semantic domain ``Bool'' which is defined as follows:
\vspace*{0.3cm}

\noindent\textbf{Domain}: Bool\\
\textbf{Operations}:
\begin{itemize}
\item \texttt{true}: Bool
\item \texttt{false}: Bool
\item and: Bool $\times$ Bool $\rightarrow$ Bool
\item or: Bool $\times$ Bool $\rightarrow$ Bool
\item not: Bool $\times$ Bool $\rightarrow$ Bool
\end{itemize}

\subsubsection{Numeral Values}
Here we consider typical domains to represent integer and float values (e.g. $\mathbb{Q}, \mathbb{N}$).

\subsubsection{Environment Values}
The domain \emph{Environment} holds the environment values of the System Architectural Model. \emph{Environment} is formalized as a tuple of domains \emph{Context} and \emph{Space}. The domain \emph{Context} is a mapping of identifiers to the environment values (\emph{Variable}, \emph{Component}, \emph{Resource}, \emph{RTEvent} and \emph{Function}), while the domain \emph{Space} models the memory space.\vspace*{0.3cm}

\noindent\textbf{Domain}: Environment\\
Environment := Context $\times$ Space\\
Context := Identifier $\rightarrow$ EnvValue
\begin{tabbing}
EnvValue := \=Variable + Component + Resource + RTEvent \\\>+ Function + AtkModel
\end{tabbing}
Space := $\mathbb{P}$(Variable)\\
Variable := n, where n $\in$ $\mathbb{N}$ represents locations
\\\textbf{Operations}:
\begin{itemize}
\item space: Environment $\rightarrow$ Space\\
space(\textless c,s\textgreater) = s
\item context: Environment $\rightarrow$ Context\\
context(\textless c,s\textgreater) = c
\item environment: Context $\times$ Space $\rightarrow$ Environment\\
environment(c,s) = \textless c, s\textgreater
\item take: Space $\rightarrow$ Identifier $\times$ Space\\
take(s) = LET x = SUCH x: x $\in$ s IN \textless x, s$\setminus$\{x\}\textgreater
\item push: Environment $\times$ Identifier $\rightarrow$ Environment
\begin{tabbing}
push(e, I) = LET \=\textless x, s'\textgreater = take(space(e)) IN \\\>environment(context(e)[I $\mapsto$ inVariable(x)], s')
\end{tabbing}
\item push: Environment $\times$ Identifier $\times$ Component $\rightarrow$ Environment
\begin{tabbing}
push(e, I, c) = LET \=\textless x, s'\textgreater = take(space(e)) IN \\\>environment(context(e)[I $\mapsto$ inComponent(c)], s')
\end{tabbing}
\item push: Environment $\times$ Identifier $\times$ AtkModel $\rightarrow$ Environment
\begin{tabbing}
push(e, I, m) = LET \=\textless x, s'\textgreater = take(space(e)) IN \\\>environment(context(e)[I $\mapsto$ inAtkModel(m)], s')
\end{tabbing}
\end{itemize}

\subsubsection{State Values}
This section defines the domain for the \emph{State} of the execution of program. A \emph{Store} is the most important part of the state and holds for every \emph{Variable} a \emph{Value}. The value can be read and modified. The \emph{Data} of the state is a tuple of a \emph{Flag} that represents the current status of the state and a \emph{Mode} to represent the current mode of execution of the state respectively component. \vspace*{0.3cm}

\noindent\textbf{Domain}: State\\
State := Store $\times$ Data\\
Store := Variable $\rightarrow$ Value\\
Data := Flag $\times$ Mode\\
Flag := \{running, ready, completed\}\\
Mode := \{normal, compromised\}
\\\textbf{Operations}:
\begin{itemize}
\item state: Store $\times$ Flag $\rightarrow$ State\\
state(s,f) = \textless s,f\textgreater
\item store: State $\rightarrow$ Store\\
store(\textless s,f\textgreater) = s
\item data: State $\rightarrow$ Data\\
data(\textless s,d\textgreater) = d
\item flag: Data $\rightarrow$ Flag\\
flag(\textless f,m\textgreater) = f
\item mode: Data $\rightarrow$ Mode\\
mode(\textless f,m\textgreater) = m
\item setFlag: State $\times$ Flag $\rightarrow$ State\\
setFlag(s, f) = LET d = \textless f, mode(data(s))\textgreater IN \textless s, d\textgreater
\item setMode: State $\times$ Mode $\rightarrow$ State\\
setMode(s, m) = LET d = \textless flag(data(s)), m\textgreater IN \textless s, d\textgreater
\item eqFlag: State $\times$ Flag $\rightarrow$ Bool\\
eqFlag(s, f) = IF equals(flag(data(s)), f) THEN \texttt{true} ELSE \texttt{false} END
\item eqMode: State $\times$ Mode $\rightarrow$ Bool\\
eqMode(s, m) = IF equals(mode(data(s)), m) THEN \texttt{true} ELSE \texttt{false} END
\item update: State $\times$ Variable $\times$ Value $\rightarrow$ State\\
update(s, var, val) = state(store(s)[var $\mapsto$ val], flag(s))
\end{itemize}

\subsubsection{Semantic Values}
\emph{Value} is a disjunctive union domain and note that the domain \emph{Value} is a recursive domain, e.g. \emph{List} is defined by \emph{Value*} as discussed in the next section.\vspace*{0.3cm}

\noindent\textbf{Domain}: Value
\begin{tabbing}
Value := \=Event + ObsEvent + RTEvent + Function + Component + Split + 
\\\>Resource + AtkModel + String + List + ... + Value$^*$
\end{tabbing}
\textbf{Operations}:
\begin{itemize}
\item equals: Value $\times$ Value $\rightarrow$ Bool
\end{itemize}

\subsubsection{Character String Values}
Character strings are defined as a semantic domain \emph{String}.

\subsubsection{Lifted Values}
The evaluation of some semantic domains might result as unsafe. To address these unsafe evaluations we lifted the domains of \emph{State} and \emph{Value} to domains $\text{State}_\bot$ and $\text{Value}_\bot$, which are disjoint sums of the basic domains and the domain $\bot$.

In order to capture different kind of events we need different semantic domain to model each of them. The three kind of events are:
\begin{enumerate}
\item \emph{Registered Events}  are the events of interest for monitoring. These events are defined by the user
at the top of the System Architectural Model by the syntactic domain ``RegModSeq'' as discussed in Appendix~\ref{sec:sam-syntax}. AWDRAT register these events for the monitoring purposes.
\item \emph{Observed Events} are the entry, exit and allowable events as defined by the syntactic domain ``Event'' of the System Architectural Model.
\item \emph{Run Time Events} are the runtime events that are generated by the monitor from the target system. These events are also called \emph{observations}.
\end{enumerate}

In the following, we give definitions of the corresponding semantic domains respectively.
\subsubsection{(Registered) Event Values}
The semantics domain \emph{Event} defines the registered events as a predicate over a sequence of input values, sequence of output values, a pre-state and a corresponding post-state as follows:
\begin{center}
Event := $\mathbb{P}$(Value$^*$ $\times$ Value$^*$ $\times$ State $\times$ State$_\bot$)
\end{center}

\subsubsection{(Observed) Event Values}
The semantics domain \emph{ObsEvent} formalizes the observed events of System Architectural Model. An \emph{ObsEvent} is defined as a predicate over a sequence of input values, a pre-state and a post-state as follows:
\begin{center}
ObsEvent := $\mathbb{P}$(Value$^*$ $\times$ State $\times$ State$_\bot$)
\end{center}
Note that the observed events do not capture output values because they just work as placeholders for runtime and registered events.

\subsubsection{(Runtime) Event Values}
The runtime events of System Architectural Model are formalized with the help of a semantic domain \emph{RTEvent}. The semantic domain \emph{RTEvent} is defined as a predicate over a sequence of input values, sequence of output values, a pre-state, post-state and event data as follows::
\begin{center}
RTEvent := $\mathbb{P}$(Value$^*$$_\bot$ $\times$ Value$^*$ $\times$ State $\times$ State$_\bot$ $\times$ EventData)
\end{center}
where\\
EventData := Tag $\times$ TimeStamp $\times$ ProcessID\\
Tag := \{entry, exit\}\\
TimeStamp := date and time of the event execution\\
ProcessID := operating system process id for the event

An \emph{EventData} captures the type of an event, time of event generation and an operating system level process id for this event. Note, process identification provides more low-level information about the event which is helpful to detect any misbehavior of the event correspondingly component.

\subsubsection{Resource Values}
The semantic domain \emph{Resource} is one of the complex domains because semantically this domain depends on the runtime behavior of an associated components as well. The semantics domain \emph{Resource} formalizes different kind of resources used by computational modules of System Architectural Model and is also defined as a predicate over a 
\begin{itemize}
\item map which is further a predicate over 
\begin{itemize}
\item a mode, 
\item its likelihood value being in normal mode, 
\item corresponding likelihood being in hacked mode and 
\item an associated vulnerability, 
\end{itemize}
\item current mode of the resource, 
\item probability of the resource being in the current mode, 
\item name of the running component associated with the resource, 
\item mode of the running associated component, 
\item a pre-state and a post-state of the program.
\end{itemize} 

The predicate \emph{Resource} is mathematically defined as:
\begin{center}
Resource := $\mathbb{P}$(ModeMap $\times$ Mode $\times$ FVal $\times$ I $\times$ Mode $\times$ State $\times$ State$_\bot$)
\end{center}
where\\
Mode := \{normal, compromised\}
\\ModeMap := $\mathbb{P}$(Mode $\times$ Fval $\times$ FVal $\times$ Vulnerability)

\subsubsection{Function Values}
The semantics domain \emph{Function} defines and formalizes a specification function of System Architectural Model and can be defined mathematically as:
\begin{center}
 $Function = \bigcup _{n \in \mathbb{N}} Function^n$
\end{center}
where
\begin{center}
 $Function^n = Value^n \rightarrow Value$
\end{center}

\subsubsection{Component Values}
The semantics domain \emph{Component} formalizes the model of the components of the target system which are specified by the corresponding behaviors in the System Architectural Model. A \emph{Component} is defined as a predicate over a structural behavior of the component, a normal behavior of the component, its corresponding compromised behavior, a pre-state and a post-state of the program as follows:
\begin{center}
Component = $\mathbb{P}$(SBehavior $\times$ NBehavior $\times$ CBehavior $\times$ State $\times$ State$_\bot$)
\end{center}
where
\begin{tabbing}
SBehavior := $\mathbb{P}$(Value$^*$ $\times$ Value$^*$ $\times$ Value$^*$ $\times$ State $\times$ State$_\bot$)\\
NBehavior = CBehavior := $\mathbb{P}$(Value$^*$ $\times$ Value$^*$ $\times$ State $\times$ State$_\bot$)
\end{tabbing}
Furthermore, a structural behavior is defined as a predicate over a sequence of input values, sequence of output value, sequence of allowable values (as a consequence of allowable events), a pre-state and a post-state of the behavior. Also,
a normal (functional) behavior and corresponding compromised behavior are defined as a predicates \emph{NBehavior} and \emph{CBehavior} over a sequence of input values, sequence of output values, a pre-state and a corresponding post-state respectively. Note that the two predicates are the valuation functions of corresponding syntactic domains.

\subsubsection{Split Values}
The semantics domain \emph{Split} formalizes the control flow behavior of a certain unit of a computational module of System Architectural Model and is defined as a predicate over a sequence of parameter values of the split as follows:
\begin{center}
Split := $\mathbb{P}$(Value$^*$)
\end{center}

\subsubsection{Attack Values}
The semantics domain \emph{AtkModel} formalizes the attack model and is defined as a predicate over an attack name, probability of the attack and the corresponding vulnerability causing the attack; the attack model is formulated as follows:
\begin{center}
AtkModel := $\mathbb{P}$(Identifier $\times$ FVal $\times$ Vulnerability)
\end{center}
These values are the result of the valuation function for the corresponding syntactic domain.

\subsection{Signatures of Valuation Functions}
A valuation function defines a mapping of a language's abstract syntax structures to its corresponding meanings (semantic algebras)~\cite{Schmidt86}. A valuation function VF for a syntax domain VF is usually formalized by a set of equations, one per alternative in the corresponding BNF Register for each syntactic domain of specification expression.

We define the result of valuation function as a predicate. In this section we first give the definitions of various relations and functions that are used in the definition of valuation functions. For example the behavioral relation (BehRelation) is defined as a predicate over an environment, a pre-state and a post-state. The corresponding relation is defined as follows:
\begin{center}
BehRelation := $\mathbb{P}$(Environment $\times$ State $\times$ State$_\bot$)
\end{center}

\subsubsection{System Architectural Model}
The valuation function for the abstract syntax domain system architectural model values of SAM is defined as follows:
\begin{center}
\textlbrackdbl SAM\textrbrackdbl : Environment $\rightarrow$ BehRelation
\end{center}

\subsubsection{Behavioral Models}
The valuation functions for abstract syntax domains of Register, structural, behavioral and split model values (RuleMod, StrMod, BehMod and SpltMod respectively) are the same and can be defined similarly; however in the following we give only the signature of valuation function for the behavioral model:
\begin{center}
\textlbrackdbl BehMod\textrbrackdbl : Environment $\rightarrow$ BehRelation
\end{center}

In the following section we define the auxiliary functions and predicates used in the formal semantics of the specification language (and associated domains).

\subsection{Auxiliary Predicates and Functions}\label{subsec:afp}
In the following subsections auxiliary functions and predicates for the use in semantics definition of sequence, binding and special expressions are defined.

\begin{itemize}
\item \begin{tabbing}
\textbf{monitors}  $\subset$ $\mathbb{N}$ $\times$ \textbf{RTEvent} $\times$ \textbf{Component} \= $\times$ \textbf{Environment}$^*$ $\times$ \textbf{Environment}$^*$
\\\>$\times$ \textbf{State}$^*$ $\times$ \textbf{State}$_\bot^*$
\\monitors(i, \textlbrackdbl rte\textrbrackdbl, \textlbrackdbl c\textrbrackdbl, e, e', s, s') $\Leftrightarrow$
\\ ( \=eqMode(s(i), ``running'') $\vee$  eqMode(s(i), ``ready'') )  $\wedge$ \textlbrackdbl c\textrbrackdbl(e(i))(e'(i), s(i), s'(i)) $\wedge$
\\\> $\exists$ oe $\in$ ObEvent: equals(rte, store(\textlbrackdbl name(rte)\textrbrackdbl)(e(i))) $\wedge$ 
\\\> IF \= entryEvent(oe, c) THEN
\\\>\> data(c, s(i), s'(i)) $\wedge$ 
\\\>\> ( preconditions(c, e(i), e'(i), s(i), s'(i), ``compromised'') $\Rightarrow$  
\\\>\>equals(s(i+1), s(i)) $\wedge$ equals(s'(i+1), s(i+1)) $\wedge$ 
\\\>\> setFlag(inState(s'(i+1)), ``compromised'') ) $\vee$
\\\>\> ( preconditions(c, e(i), e'(i), s(i), s'(i), ``normal'')
\\\>\> $\Rightarrow$ setMode(s(i), ``running'') $\wedge$ 
\\\>\> LET \= cseq = components(c) IN
\\\>\>\> equals(s(i+1), s'(i)) $\wedge$ equals(e(i+1), e'(i)) $\wedge$
\\\>\>\>$\forall$ \= c$_1$ $\in$ cseq, rte$_1$ $\in$ RTEvent: 
\\\>\>\>\> arrives(rte$_1$, s(i+1)) $\wedge$ 
\\\>\>\>\> monitor(i+1, rte$_1$, c$_1$, e(i+1), e'(i+1), s(i+1), s'(i+1))
\\\>\> END )
\\\> ELSE IF \= exitEvent(oe, c) THEN
\\\>\> data(c, s(i), s'(i)) $\wedge$ eqMode(inState(s'(i)), ``completed'') $\wedge$ 
\\\>\>postco\=nditions(c, e(i), e'(i), s(i), s'(i), ``normal'') 
\\\>\>\> $\Rightarrow$ \= equals(s(i+1), s'(i)) $\wedge$ equals(e(i+1), e'(i)) $\wedge$ 
\\\>\>\>\> setMode(inState(s'(i+1), ``completed'')
\\\> ELSE IF \= allowableEvent(oe, c) THEN
\\\>\> equals(s(i+1), s'(i)) $\wedge$ equals(e(i+1), e'(i))
\\\> ELSE \= 
\\\>\>equals(s(i+1), s(i)) $\wedge$ equals(s'(i+1), s(i+1)) $\wedge$ 
\\\>\> setFlag(inState(s'(i+1)), ``compromised'')
\\\> END
\end{tabbing}
The predicate ``monitors'' captures the core semantics of the monitor which is defined as a relation on
\begin{itemize}
\item number of observation $i$ with respect to iteration of a component,
\item an observation (runtime event) $rte$,
\item corresponding component $c$ under observation,
\item a sequence of pre-environments $e$,
\item a sequence of post-environments $e'$,
\item a sequence of pre-states $s$ and
\item a sequence of post-states $s'$.
\end{itemize}
The predicate $monitors$ is defined such that, at any arbitrary observation if the current execution state $s(i)$ of component $c$ is ``ready'' or ``running'' and behavior of component $c$ has been evaluated and there is a \emph{prediction} $oe$ which is semantically equal to an \emph{observation} $rte$ and any of the following
can happen:
\begin{itemize}
\item either the \emph{prediction} respectively \emph{observation} is  an entry event of the component $c$, then it waits until the complete data for the component $c$ arrives, if so, then
\begin{itemize}
\item either preconditions of ``normal'' behavior of the component hold; if so then, the subnetwork of the component is initiated and the components in the subnetwork are monitored iteratively with the corresponding arrival of the \emph{observation}
\item or preconditions of ``compromised''  behavior of the component hold, in this case the state is marked to ``compromised'' and returns
\end{itemize}
\item or the \emph{observation} is an exit event and after the completion of data arrival the postconditions hold and the resulting state is marked as ``completed''
\item or the \emph{observation} is an allowable event and just continues the execution
\item or the \emph{observation} is an unexpected event (or any of the above does not hold), then the state is marked as ``compromised'' and returns.
\end{itemize}
The predicate \emph{monitors} is used later in the semantics of the Execution Monitor.

\item \textbf{entryEvent $\subset$ ObEvent $\times$ Component}: returns \texttt{true} only if the given event is in a set of entry events of the given component.
\item \textbf{exitEvent $\subset$ ObEvent $\times$ Component}: returns \texttt{true} only if the given event is in a set of exit events of the given component.
\item \textbf{allowableEvent $\subset$ ObEvent $\times$ Component}: returns \texttt{true} only if the given event is in a set of allowable events of the given component.
\item \textbf{data $\subset$ Component $\times$ State $\times$ State$_\bot$}: returns \texttt{true} only if all the data for the given component is received by transforming a given pre-state (former) into a corresponding given post-state (latter).
\item \textbf{arrives $\subset$ RTEvent $\times$ State}: returns \texttt{true} only if the given runtime event (observation) arrives in a given state.
\item \begin{tabbing}\textbf{preconditions} $\subset$ \textbf{Component} \=$\times$ \textbf{Environment} $\times$ \textbf{Environment} \\\>$\times$ \textbf{State} $\times$ \textbf{State}$_\bot$ $\times$ \textbf{Mode}
\end{tabbing} returns \texttt{true} only if all the preconditions of the given component hold in a given pair of pre- and post-environments, a pair of pre- and post-states and in a given mode.
\item \begin{tabbing}\textbf{postconditions} $\subset$ \textbf{Component} \=$\times$ \textbf{Environment} $\times$ \textbf{Environment} \\\>$\times$ \textbf{State} $\times$ \textbf{State}$_\bot$ $\times$ \textbf{Mode}
\end{tabbing}
returns \texttt{true} only if all the postconditions of the given component hold in a given pair of pre- and post-environments, a pair of pre- and post-states and in a given mode.
\item \textbf{startup $\subset$ State $\times$ Target\_System}: returns \texttt{true} only if the given state is an initial state of the execution of the given target system.
\item \textbf{isTop $\subset$ Component $\times$ (Environment $\rightarrow$ BehRelation)}: returns \texttt{true} only if the given component is a top level component of the given semantics of a System Architectural Model.
\item \textbf{enableDiagnosis: Environment} $\rightarrow$ $\mathbb{P}$(\textbf{State $\times$ Value}): results in a given recovered state and a boolean value (\texttt{true}, if recovered safely, \texttt{false} otherwise) from a given environment.
\item \begin{tabbing}\textbf{respectsOrder} $\subset$ \textbf{Identifier\_Sequence} \=$\times$ \textbf{Identifier\_Sequence} 
\end{tabbing}
returns \texttt{true} only if the identifiers in the latter sequence has the same order as the identifiers in the
former sequence.
\item \begin{tabbing}\textbf{buildEnv} $\subset$ \textbf{Environment} \=$\times$ \textbf{List*} $\times$ \textbf{List*} $\times$ \textbf{List*} $\times$ \textbf{List*} \\\>$\rightarrow$ \textbf{Environment}
\end{tabbing}
builds the resulting environment by updating the given environment with given sequences of values of
\begin{itemize}
\item resources
\item resource mappings
\item model mappings and
\item vulnerabilities
\end{itemize}
respectively.
\end{itemize}

\subsection{Definition of Valuation Functions}
In this section we give the definition of the formal semantics of the interesting syntactic domains (and associated domains) of the specification language, e.g. system architectural model, Register mode, behavioral model and split model. The semantics of other domains of the specification language are very simple and can be easily rehearsed.

\subsubsection{System Architectural Model}
The System Architectural Model is give by the syntactic domain \emph{SAM} such that
\begin{itemize}
\item the Register model (syntactic domain \emph{RegModSeq}) is defined at the top of the System Architectural Model 
which gives the registered events to be monitored at runtime
\item followed by 
\begin{itemize}
\item a hierarchical structural behavior (syntactic domain \emph{StrModSeq}) of components,
\item a normal respectively compromised behavior (syntactic domain \emph{BehModSeq}) and
\item corresponding split behaviors (syntactic domain \emph{SplModSeq}) occurring in any of the structural behavior of the components.
\end{itemize}
\end{itemize}
For further details on the syntax of the model, please see Appendix~\ref{sec:sam-syntax}.

Semantically, an overall (system architectural) model holds (\texttt{true}) in a given environment $e$ such that it produces a new environment $e'$ and a post-state $e'$ when executed in a pre-state $s$ as defined below.

\begin{tabbing}
\textlbrackdbl \=SAM\textrbrackdbl (e)(e', s, s') $\Leftrightarrow$ 
\\\> $\forall$ \=$e_1, e_2, e_3 \in$ Environment, $s_1, s_2, s_3 \in $ State: \\\>\> \textlbrackdbl RegModSeq\textrbrackdbl (e)(e$_1$, s, inState$_\bot$(s$_1$)) $\wedge$ \textlbrackdbl StrModSeq\textrbrackdbl (e$_1$)(e$_2$, s$_1$, inState$_\bot$(s$_2$)) $\wedge$
\\\>\>\textlbrackdbl BehModSeq\textrbrackdbl (e$_2$) (e$_3$, s$_2$, inState$_\bot$(s$_3$))
$\wedge$ \textlbrackdbl SpltModSeq\textrbrackdbl (e$_3$)(e', s$_3$, s')
\end{tabbing}
In detail, the semantics of the System Architectural Model \emph{SAM} holds in a given environment $e$ resulting in an environment $e'$ by transforming a pre-state $s$ into post-state $s'$ and
\begin{itemize}
\item the evaluation of the registered events in a given environment $e$ results in environment $e_1$ transforming a pre-state $s$ into a post-state $s_1$ and in principle
\begin{itemize}
\item the structural behavior of components hold in environment $e_1$ (with some auxiliary transformations) and
\item the functional behavior of components hold in environment $e_2$ (with some auxiliary transformations) and finally
\item the split behavior of components hold in $e_3$ resulting in given environment $s'$ and transforming a pre-state $s_3$ into a given post-state $s'$.
\end{itemize}
\end{itemize}

In the following, first we define the semantics of unit elements \emph{RegMod}, \emph{StrMod}, \emph{BehMod} and \emph{SplModSeq}
and then define corresponding sequence domains \emph{RegModSeq}, \emph{StrModSeq}, \emph{BehModSeq} and \emph{SplModSeq} respectively.

\subsubsection{Register Model}
The syntactic domain (RegMod) defines a registered event as follows:
\begin{tabbing}
RegMod ::= \textbf{register-event} \='EvntName JavClaName JavMetName '(JavParamSeq)
\\\>\=[ \textbf{:static} ObjName ] 
\\\>\>[ \textbf{:output-type} JavParam ] 
\\\>\>[ \textbf{:bypass} ObjNameStr ]
\\\>\>[ \textbf{:}EvntName ObjName]
\end{tabbing}
Though the domain represents language independent event registration, in this document we focus only on the the Java based target system. The syntactic phrase \emph{RegMod} states that a registered event can be represented by a name (EvntName) whose source is a Java method (JavMetName) with parameters (JavParmSeq) of corresponding class (JavClaName). The other sub-clauses introduce further characterization of the method, e.g. the clause \textbf{:output-type} represents the return type of the method.

A monitoring machinery of Architectural Differencer of the middleware AWDRAT is based on these registered events.

In the following we define the semantics of a registered event such that the evaluation of a registered event in a given environment $e$ results in an environment $e'$ transforming a pre-state $s$ into a post-state $s'$.
\begin{tabbing}
\textlbrackdbl \=RegMod\textrbrackdbl (e)(e', s, s') $\Leftrightarrow$ 
\\\> $\forall$ \=e$_j$ $\in$ JTypeEnvironment, s$_j$, s$_j$' $\in$ JState: 
\\\>\> typeCheck(JavClaName)(e$_j$)(s$_j$, s$_j'$) $\wedge$ equals(s, s$_j$) $\wedge$ equals(e, e$_j$)\\\>\>$\wedge$ ($\exists$ \=p $\in$ JProcedure: equals(p(valseq, val), store(s$_j$')(\textlbrackdbl JavMetName\textrbrackdbl(e$_j$)))
\\\>\>\> $\wedge$ equals(valseq, store(s$_j$')(\textlbrackdbl JavParamSeq\textrbrackdbl(e$_j$))) 
\\\>\>\> $\wedge$ equals(val, store(s$_j$')(\textlbrackdbl JavParam\textrbrackdbl(e$_j$))))
\\\>\> $\wedge$ isStatic(...) $\wedge$ byPass(...) $\wedge$ otherEvents(...)
\\\>\> $\wedge$ e' = push(e, EvntName) 
\\\>\> $\wedge$ \=LET \=ev(valseq, val, s, s') $\in$ Event IN  
\\\>\>\>\> s' = update(s, \textlbrackdbl EvntName\textrbrackdbl(e'), ev)
\\\>\>\> END
\end{tabbing}
In detail, semantically, the Java class (JavClaName) is well-defined respectively type checked in an arbitrary environment $e_j$ transforming an arbitrary pre-state $s_j$ into a corresponding arbitrary post-state $s_j$' while $e_j$ and $s_j$ are semantically equivalent to $s$ and $e$ respectively and
\begin{itemize}
\item there is some Java procedure $p(valseq, val)$ which we get by evaluating ``JavMetName'' with given environment $e_j$ such that the sequence of input values $valseq$ equals the evaluation of ``JavParamSeq'' in given environment $e_j$ and the return value $val$ of procedure $p$ equals the evaluation of ``JavParam'' in environment $e_j$ and
\item finally we get $e'$ by pushing ``EvntName'' in given environment $e$ and
\item $s'$ is produces by updating the value $ev$ for an identifier ``EvntName'' in the given pre-state $s$.
\end{itemize}

\subsubsection{Register Model Sequence}
The syntax of the syntactic domain RegModSeq is defined as follows:\vspace*{0.3cm}

\noindent RegModSeq := EMPTY $|$ (RegMod) RegModSeq

Semantically, when an EMPTY sub-phrase is evaluated in a given environment $e$ then simply the resulting environment $e'$ equals $e$ and a post-state $s'$ equals the given pre-state $s$ as defined below:
\subsubsection*{Case: EMPTY}
\begin{tabbing}
\textlbrackdbl \=EMPTY\textrbrackdbl (e)(e', s, s') $\Leftrightarrow$ e' = e $\wedge$ s' = inState$_\bot$(s)
\end{tabbing}

While in the second alternate of the domain ``RegModSeq'', first semantics of the phrase ``RegMod'' in a given environment $e$ produce an environment $e''$ transforming a pre-state $s$ into a post-state $s''$, then the evaluation of the phrase ``RegModSeq'' in environment $e''$ results in a given environment $e'$ and transforms the pre-state $s''$ into a given post-state $s'$. The semantics of the second alternative is formalized as follows:

\subsubsection*{Case: (RegMod) RegModSeq}
\begin{tabbing}
\textlbrackdbl \=(RegMod) RegModSeq\textrbrackdbl (e)(e', s, s') $\Leftrightarrow$ 
\\\> $\forall$ \= e'' $\in$ Environment, s'' $\in$ State: 
\\\>\>\textlbrackdbl RegMod\textrbrackdbl (e)(e'', s, inState$_\bot$(s'')) $\wedge$ \textlbrackdbl RegModSeq\textrbrackdbl (e'')(e', s'', s')
\end{tabbing}

\subsubsection{Structural Model}
The structural behavior of the system is defined by the syntactic phrase ``StrMod'' which represents a corresponding hierarchical model of the components. The syntax for the overall structural behavior of the component ``CompName'' is defined by the syntactic phrase ``StrMod'' where different clauses define three logical parts of the behavior as follows:
\begin{enumerate}
\item \emph{signals} specify global control behavior of the component, e.g. 
\begin{itemize}
\item the clauses \textbf{:entry-events} and \textbf{:exit-events} models the entry and exit events of the component respectively and
\item the other allowable events (while execution of the component) are modeled with the clause \textbf{:allowable-events}
\end{itemize}
\item \emph{signature} of the component consists of
\begin{itemize}
\item the sequences of objects for the clauses \textbf{:inputs} and \textbf{:outputs} respectively
\end{itemize}
\item \emph{body} of the component is modeled as a sub-network which involves different components as represented by the \textbf{:components} clause. These components are connected through various nodes and links as follows:
\begin{enumerate}
\item the control flows \textbf{:controlflows} which further have corresponding splits \textbf{:splits} and joins \textbf{:joins} and
\item the propagation of data among the components (via control flows) is represented by the clause \textbf{:dataflows}.
\item while the execution of the body, various computing resources \textbf{:resources} (each with a name, its type and its probabilities of being in normal and hacked modes respectively) are involved which further requires
\item the resource mappings \textbf{:resource-mappings} (where each resource is mapped to a component that uses it) in addition to
\item the model mappings \textbf{:model-mappings} (where the conditional probability between the compromises and misbehaviors for each of the component is given) and
\item the vulnerabilities \textbf{:vulnerabilities} such that each resource is mapped to a corresponding (possible) vulnerability (which is assumed to be defined as the part of an attack plan that is beyond the scope of this document).
\end{enumerate}
\end{enumerate}

The syntactic domain of for the structural behavioral model (StrMod) is defined as follows:

\begin{tabbing}
StrMod ::= \textbf{define-ensemble} \=CompName
\\\> \textbf{:entry-events} \textbf{:auto} $|$ (EvntSeq$_1$)
\\\> \textbf{:exit-events} (EvntSeq$_2$)
\\\> \textbf{:allowable-events} (EvntSeq$_3$)
\\\> \textbf{:inputs} (ObjNameSeq$_1$)
\\\> \textbf{:outputs} (ObjNameSeq$_2$)
\\\> \textbf{:components} (CompSeq)
\\\> \textbf{:controlflows} (CtrlFlowSeq)
\\\> \textbf{:splits} (SpltCFSeq)
\\\> \textbf{:joins} (JoinCFSeq)
\\\> \textbf{:dataflows} (DataFlowSeq)
\\\> \textbf{:resources} (ResSeq)
\\\> \textbf{:resource-mapping} (ResMapSeq)
\\\> \textbf{:model-mappings} (ModMapSeq)
\\\> \textbf{:vulnerabilities} (VulnrabltySeq)
\end{tabbing}

The semantics of the structural behavioral model in a given environment $e$ results in an environment $e'$ transforming a pre-state $s$ into a post-state $s'$ as defined below:

\begin{tabbing}
\textlbrackdbl \=StrMod\textrbrackdbl (e)(e', s, s') $\Leftrightarrow$
\\\> $\forall$ \=$e, e_1, e_2, e_3, e_4, e_5, e_6, e_7, e_8 \in$ Environment, $s, s_1, s_2, s_3, s_4, s_5, s_6, s_7, s_8 \in $ State, 
\\\>\> oeseq, oeseq$_1$, aeseq $\in$ ObsEvent*, anameseq, enameseq, enameseq$_1$ $\in$ EvntNameSeq:
\\\>\> ( eqFlag(s, ``running'') $\wedge$  
\\\>\> \textlbrackdbl EvntSeq$_3$\textrbrackdbl(e)(e$_1$, s, inState$_\bot$(s$_1$), enameseq, oeseq) $\wedge$ 
\\\>\>$\forall$ \= ename $\in$ enameseq: 
\\\>\>\>$\exists$ se $\in$ Event, rte $\in$ RTEvent, oe $\in$ oeseq: 
\\\>\>\> IF \= equals(se, oe) THEN 
\\\>\>\>\>LET \=rte = store(s$_1$)(ename) IN 
\\\>\>\>\>\> IF \=equals(rte, se) THEN
\\\>\>\>\>\> \>\texttt{true} 
\\\>\>\>\>\> ELSE 
\\\>\>\>\>\>\> s$_1$ = enableDiagnosis(e$_1$)(s$_1$, inBool(\texttt{true}))
\\\>\>\>\>\> END
\\\>\>\>\> END
\\\>\>\> ELSE
\\\>\>\>\> s$_1$ = enableDiagnosis(e$_1$)(s$_1$, inBool(\texttt{true}))
\\\>\>\> END )
\\\>\> $\vee$ 
\\\>\> ( eqFlag(s, ``running'') $\vee$ eqFlag(s, ``ready'')  $\wedge$ 
\\\>\> \textlbrackdbl EvntSeq$_1$\textrbrackdbl(e)(e$_1$, s, inState$_\bot$(s$_1$), enameseq, oeseq) $\wedge$ 
\\\>\>$\forall$ \= ename $\in$ enameseq, oe $\in$ oeseq: 
\\\>\>\>$\exists$ se $\in$ Event, rte $\in$ RTEvent: 
\\\>\>\>equals(se, store(s$_1$)(ename)) $\wedge$ equals(se[1], oe[1]) $\wedge$ 
\\\>\>\> LET \=rte = store(s$_1$)(ename) IN 
\\\>\>\>\>IF \=equals(rte[5][1], ``entry'' ) THEN 
\\\>\>\>\>\>equals(rte[1], se[1])
\\\>\>\>\>ELSE equals(rte[2], se[2])
\\\>\>\>\>END 
\\\>\>\>END $\wedge$ 
\\\>\>\> $\forall$ \= inseq $\in$ Value$^*$, c $\in$ Component: 
\\\>\>\>\>\textlbrackdbl ObjNameSeq$_1$\textrbrackdbl (e$_1$)(inState$_\bot$(s$_1$), inseq) $\wedge$ \textlbrackdbl CompName\textrbrackdbl(e$_1$)(inValue(c)) $\wedge$
\\\>\>\>\> IF \= equals(c[2][1], inseq) THEN 
\\\>\>\>\>\> eqMode(s$_1$, ``normal'') 
\\\>\>\>\> ELSE  
\\\>\>\>\>\> s$_1$ = enableDiagnosis(e$_1$)(s$_1$, inBool(\texttt{true})) 
\\\>\>\>\> END $\wedge$ 
\\\>\>\>\> IF equals(c[3][1], inseq) THEN 
\\\>\>\>\>\> eqMode(s$_1$, ``compromised'') $\wedge$ s$_1$ = enableDiagnosis(e$_1$)(s$_1$, inBool(\texttt{true})) 
\\\>\>\>\> ELSE \texttt{true}
\\\>\>\>\> END )
\\\>\> $\Rightarrow$ \= eqFlag(s$_1$, ``running'') $\wedge$ 
\\\>\>\> $\forall$ compseq $\in$ Component$^*$: \textlbrackdbl CompSeq\textrbrackdbl(e$_2$)(e$_3$, s$_2$, inState$_\bot$(s$_3$), compseq) $\wedge$
\\\>\>\> $\forall$ \= rmseq, crmapseq, cpmapseq, vbltyseq $\in$ List$^*$: 
\\\>\>\>\>\textlbrackdbl ResSeq\textrbrackdbl (e$_3$)(s$_3$, inState$_\bot$(s$_4$), rmseq) $\wedge$
\\\>\>\>\> \textlbrackdbl ResMapSeq\textrbrackdbl (e$_3$)(s$_3$, inState$_\bot$($s_4$), crmapseq) $\wedge$
\\\>\>\>\> \textlbrackdbl ModMapSeq\textrbrackdbl (e$_3$)(s$_3$, inState$_\bot$($s_4$), cpmapseq) $\wedge$
\\\>\>\>\> \textlbrackdbl VulnrabltySeq\textrbrackdbl (e$_3$)(s$_3$, inState$_\bot$($s_4$), vbltyseq) $\wedge$ 
\\\>\>\>\> e$_4$ = buildEnv(e$_3$, rmseq, crmapseq, cpmapseq, vbltyseq) $\wedge$
\\\>\>\>\>\textlbrackdbl CtrlFlowSeq\textrbrackdbl (e$_4$)(e$_5$, s$_4$, inState$_\bot$(s$_5$)) $\wedge$ 
\\\>\>\>\> \textlbrackdbl SpltCFSeq\textrbrackdbl (e$_5$)(e$_6$, s$_5$, inState$_\bot$(s$_6$)) $\wedge$
\\\>\>\>\>\textlbrackdbl JoinCFSeq\textrbrackdbl (e$_6$)(e$_7$, s$_6$, inState$_\bot$(s$_7$)) $\wedge$ 
\\\>\>\>\> \textlbrackdbl DataFlowSeq\textrbrackdbl (e$_7$)(e$_8$, s$_7$, inState$_\bot$(s$_8$)) $\wedge$  
\\\>\>\>\> \textlbrackdbl EvntSeq$_2$\textrbrackdbl(e$_8$)(e$_9$, s$_8$, s', enameseq$_1$, oeseq$_1$) $\wedge$
\\\>\>\>\> $\forall$ \= ename $\in$ enameseq, oe $\in$ oeseq: 
\\\>\>\>\>\>$\exists$ se $\in$ Event, rte $\in$ RTEvent: 
\\\>\>\>\>\>equals(se, store(inState(s'))(ename)) $\wedge$ equals(se[1], oe[1]) $\wedge$ 
\\\>\>\>\>\> LET \=rte = store(inState(s'))(ename) IN 
\\\>\>\>\>\>\>IF \=equals(rte[5][1], ``entry'' ) THEN 
\\\>\>\>\>\>\>\>equals(rte[1], se[1])
\\\>\>\>\>\>\>ELSE equals(rte[2], se[2])
\\\>\>\>\>\>\>END 
\\\>\>\>\>\>END
\\\>\>\>\>\> $\Rightarrow$\= 
\\\>\>\>\>\>\> $\forall$ \= outseq $\in$ Value$^*$, c $\in$ Component: 
\\\>\>\>\>\>\>\>\textlbrackdbl ObjNameSeq$_2$\textrbrackdbl (e$_9$)(s', outseq) $\wedge$ \textlbrackdbl CompName\textrbrackdbl(e$_9$)(inValue(c)) $\wedge$
\\\>\>\>\>\>\>\> ( ( IF \= equals(c[2][2], outseq) THEN
\\\>\>\>\>\>\>\>\> eqMode(inState(s'), ``normal'')
\\\>\>\>\>\>\>\> ELSE
\\\>\>\>\>\>\>\>\> s' = enableDiagnosis(e$_9$)(inState(s'), inBool(\texttt{true}))
\\\>\>\>\>\>\>\> END ) $\vee$ 
\\\>\>\>\>\>\>\> ( IF \= equals(c[3][2], outseq) THEN
\\\>\>\>\>\>\>\>\> eqMode(inState(s'), ``compromised'') $\wedge$ 
\\\>\>\>\>\>\>\>\> s' = enableDiagnosis(e$_9$)(inState(s'), inBool(\texttt{true}))
\\\>\>\>\>\>\>\> END ) ) $\wedge$ 
\\\>\>\>\>\>\>\> eqMode(inState(s'), ``normal'') $\wedge$
\\\>\>\>\>\>\>\> eqFlag(inState(s'), ``completed'') $\wedge$
\\\>\>\>\>\>\>\> LET \= sbeh = \textless inseq, outseq, s, s'\textgreater, nbeh = c[2], cbeh = c[3] IN 
\\\>\>\>\>\>\>\>\> e' = push(e$_9$\=, store(inState(s'))(\textlbrackdbl CompName\textrbrackdbl(e$_9$))
\\\>\>\>\>\>\>\>\>\>, c(sbeh, nbeh, cbeh, s, s'))
\\\>\>\>\>\>\>\> END
\end{tabbing}

In general, the semantics is defined as a big logical implication, where the premise is a disjunction of two formulas
as explained below:
\begin{enumerate}
\item either the current state $s$ of the component is ``running'' and it receives allowable events ``EvntSeq$_3$'' and 
for every event $oe$ in the allowable event sequence $oeseq$ there is a corresponding equivalent registered event $se$ for which we receive an equivalent runtime event $rte$ such that $rte$ is one of the under observation (legal) event $se$, if not then the runtime event is a result of the misbehavior of the component so the diagnosis component of AWDRAT is activated by calling ``enableDiagnosis(...)'' which successfully recovers the compromised state $s_1$
\item or the current state $s$ of the component is either ``running'' or ``ready'' and it receives the entry events ``EvntSeq$_1$'' (evaluating to $oeseq$) and for every event $oe$ in the sequence of entry events $oeseq$ there is a corresponding registered event $se$ and the received runtime event $rte$ (equals $se$ depending on its type ``entry'' or ``exit'') is the monitored event and
\begin{itemize}
\item if the sequence of input values $inseq$ satisfies the pre-conditions of the ``normal'' behavior (c[2][1])) of the component (c) then the resulting state $s_1$ is in ``normal'' mode
\item otherwise (when pre-conditions are not satisfied) then the diagnosis component is activated which recovers the compromised state $s_1$ and
\item if the sequence of input values $inseq$ satisfies the pre-conditions of the already ``compromised'' behavior (c[3][1]) of the component (c) then the resulting state $s_1$ is ``compromised'' state and we restore it by enabling diagnostic engine
\item otherwise the system is safe \texttt{true} to start executing component respectively \emph{body}/sub-network.
\end{itemize}
\end{enumerate}

Semantically, if any of the above two holds then the
\begin{itemize}
\item the current state $s_1$ is ``running'' and the components of the sub-network evaluate to $compseq$ and
\item a new environment $e_4$ is constructed based on the evaluation of the resources, resource mappings, model mappings and vulnerabilities of the sub-network to $rmseq$, $crmapseq$, $cpmapseq$, and $vbltyseq$ respectively (such
that all the trust model of the components is known before the actual execution of the \emph{body} starts) in which
\item the execution blocks are evaluated (such that the evaluation of the control flows their respective splits and joins and associated data flows results in an environment $e_8$ and a post-state $s_8$) to complete the executional behavior and
\item once all the sub-network is executed (recursively), then the receiving exit events (EvntSeq$_2$) evaluate to $oeseq_1$ and if for every event $oe$ in $oeseq_1$ there is an equivalent registered event $se$ and a runtime event $rte$ then
\begin{itemize}
\item either the sequence of output values $outseq$ satisfies the post-conditions of the ``normal'' behavior (c[2][2]) of the component (c) then the post-state $s'$ is in ``normal'' mode otherwise the diagnosis component restores the
post-state $s'$ and
\item or the sequence of output values $outseq$ satisfies the post-conditions of the misbehavior (c[3][2]) of the component (c) then the post-state must be in ``compromised'' mode and corresponding diagnosis component is enabled to recover
the state back and
\end{itemize}
\item the given and transformed final post-state $s'$ must be in ``normal'' mode with ``completed'' flag and
\item finally the resulting environment $e'$ is build with the evaluated behavior of the component of the current
component.
\end{itemize}

\subsubsection{Structural Model Sequence}
The syntactic domain StrModSeq is:\vspace*{0.3cm}

\noindent StrModSeq := EMPTY $|$ (StrMod) StrModSeq

\subsubsection*{Case: EMPTY}
\begin{tabbing}
\textlbrackdbl \=EMPTY\textrbrackdbl (e)(e', s, s') $\Leftrightarrow$ e' = e $\wedge$ s' = inState$_\bot$(s)
\end{tabbing}
\subsubsection*{Case: (StrMod) StrModSeq}
\begin{tabbing}
\textlbrackdbl \=(StrMod) StrModSeq\textrbrackdbl (e)(e', s, s') $\Leftrightarrow$ 
\\\> $\forall$ \= e'' $\in$ Environment, s'' $\in$ State: 
\\\>\>\textlbrackdbl StrMod\textrbrackdbl (e)(e'', s, inState$_\bot$(s'')) $\wedge$ \textlbrackdbl StrModSeq\textrbrackdbl (e'')(e', s'', s')
\end{tabbing}
The semantics of the domain ``StrModSeq'' of structural model sequence are similar to the semantics of the domain ``RegModSeq'' as discussed above in the corresponding section. Similarly, the semantics of the syntactic domains of ``BehModSeq'' and ``SplModSeq'' can be exercised which are discussed in the corresponding sections later in this document.

\subsubsection{Behavioral Model}
The behavioral model represents the functional behavior of a component, which can be either ``normal'' or known ``compromised'' one. The functional behavior of the component ``CompName'' consists of the following elements:
\begin{enumerate}
\item the inputs of the component as given by the clause \textbf{:inputs},
\item the outputs of the component as represented by the corresponding clause \textbf{:outputs},
\item the allowable events \textbf{:allowable-events} represents the auxiliary communication of the component,
\item the pre-conditions of the component are specified in the clause \textbf{:prerequisites} while
\item the corresponding post-conditions are specified by the \textbf{:postconditions} clause.
\end{enumerate}
Note that the ``compromised'' behavior is used to model already known misbehaviors of the component (e.g.
some attack) and needs corresponding diagnosis which in this case is already known.

The syntactic domain ``BehMod'' for the behavioral model is defined as follows: 
\begin{tabbing}
BehMod ::= \textbf{defbehavior-model} \=(CompName \textbf{normal} $|$ \textbf{compromised})
\\\> \textbf{:inputs} (ObjNameSeq$_1$)
\\\> \textbf{:outputs} (ObjNameSeq$_2$)
\\\> \textbf{:allowable-events} (EvntSeq)
\\\> \textbf{:prerequisites} (BehCondSeq$_1$)
\\\> \textbf{:postconditions} (BehCondSeq$_2$)
\end{tabbing}

Semantically, normal and compromised behavioral models results in modifying the corresponding elements of the environment value ``Component'' as defined below:
\begin{tabbing}
\textlbrackdbl \=BehMod\textrbrackdbl (e)(e', s, s') $\Leftrightarrow$ 
\\\> $\forall$ \=e$_1$ $\in$ Environment, nseq $\in$ EvntNameSeq, eseq $\in$ ObsEvent*, inseq, outseq $\in$ Value$^*$: \\\>\> \textlbrackdbl ObjNameSeq$_1$\textrbrackdbl (e)(inState$_\bot$(s), inseq) $\wedge$ \textlbrackdbl BehCondSeq$_1$\textrbrackdbl (e) (inState$_\bot$(s)) $\wedge$ 
\\\>\> \textlbrackdbl EvntSeq\textrbrackdbl (e) (e$_1$, s, s', nseq, eseq)
\\\>\> \textlbrackdbl ObjNameSeq$_2$\textrbrackdbl (e$_1$)(s', outseq) $\wedge$ \textlbrackdbl BehCondSeq$_2$\textrbrackdbl (e$_1$) (s') $\wedge$
\\\>\> $\exists$ \= c $\in$ Component: \textlbrackdbl CompName\textrbrackdbl(e$_1$)(inValue(c)) $\wedge$
\\\>\> IF \=eqMode(inState$_\bot$(s'), ``normal'') THEN
\\\>\>\> LET \= sbeh = c[1], nbeh = \textless inseq, outseq, s, s'\textgreater, cbeh = c[3] IN 
\\\>\>\>\> e' = push(e$_1$\=, store(inState(s'))(\textlbrackdbl CompName\textrbrackdbl(e$_1$))
\\\>\>\>\>\>, c(sbeh, nbeh, cbeh, s, s'))
\\\>\>\> END
\\\>\> ELSE
\\\>\>\> LET \= sbeh = c[1], nbeh = c[2], cbeh = \textless inseq, outseq, s, s'\textgreater~ IN 
\\\>\>\>\> e' = push(e$_1$\=, store(inState(s'))(\textlbrackdbl CompName\textrbrackdbl(e$_1$))
\\\>\>\>\>\>, c(sbeh, nbeh, cbeh, s, s'))
\\\>\>\> END
\\\>\> END
\end{tabbing}

In detail, if the semantics of of syntactic domain ``BehMod'' holds in a given environment $e$ resulting in environment $e'$ and transforming a pre-state $s$ into corresponding post-state $s'$ then
\begin{itemize}
\item the inputs ``ObjNameSeq$_1$'' evaluates to a sequence of values $inseq$ in a given environment $e$ and a given state $s$ which satisfies the corresponding pre-conditions ``BehCondSeq$_1$'' in the same $e$ and $s$ and
\item the allowable events happens whose evaluation results in new environment $e_1$ and given post-state $s'$ with some auxiliary sequences $nseq$ and $eseq$ and
\item the outputs ``ObjNameSeq$_2$'' evaluates to a sequence of values $outseq$ in an environment $e_1$ and given post-state $s'$ which satisfies the corresponding post-conditions ``BehCondSeq$_2$'' in the same environment $e_1$ and state $s'$ and the given environment $e'$ can be constructed such that
\begin{itemize}
\item if the post-state is ``normal'' then $e'$ is an update to the normal behavior ``nbeh'' of the component ``CompName'' in environment $e_1$
\item otherwise $e'$ is an update to the compromised behavior ``cbeh'' of the component.
\end{itemize}
\end{itemize}
In the construction of the environment $e'$ the rest of the semantics of the component do not change as represented in the corresponding LET-IN constructs. 

\subsubsection{Behavioral Model Sequence}
The syntactic domain BehModSeq is:\vspace*{0.3cm}

\noindent BehModSeq := EMPTY $|$ (BehMod) BehModSeq

\subsubsection*{Case: EMPTY}
\begin{tabbing}
\textlbrackdbl \=EMPTY\textrbrackdbl (e)(e', s, s') $\Leftrightarrow$ e' = e $\wedge$ s' = inState$_\bot$(s)
\end{tabbing}
\subsubsection*{Case: (BehMod) BehModSeq}
\begin{tabbing}
\textlbrackdbl \=(BehMod) BehModSeq\textrbrackdbl (e)(e', s, s') $\Leftrightarrow$ 
\\\> $\forall$ \= e'' $\in$ Environment, s'' $\in$ State: 
\\\>\>\textlbrackdbl BehMod\textrbrackdbl (e)(e'', s, inState$_\bot$(s'')) $\wedge$ \textlbrackdbl BehModSeq\textrbrackdbl (e'')(e', s'', s')
\end{tabbing}

\subsubsection{Split Model}
Though the splits of control flows are declared in the ``StrBeh'' domain but their corresponding definitions are given with the help of the domain ``SplMod'' which consists of its
\begin{itemize}
\item name ``SpltModName'',
\item required sequence of parameters ``SpltParamSeq'' which are used by the various branches of the split as defined in
\item the split condition branches ``SpltCondSeq''.
\end{itemize}

The syntax of the domain ``SplMod'' is given as follow: \vspace*{0.3cm}

\noindent SplMod ::= \textbf{defsplit} SpltModName\textbf{?} (SpltParamSeq) SpltCondSeq)\vspace*{0.3cm}

If the semantics of the split model ``SplMod'' in a given environment $e$ results in environment $e'$ and transforms a pre-state $s$ into post-state $s'$ then
\begin{itemize}
\item first the parameters are evaluated in a given environment $e$ which results in an environment $e_1$ and sequence of values $vseq$ transforming a given pre-state $s$ into post-state $s_1$ and
\item the split conditions ``SpltCondSeq'' hold in environment $e_1$ producing environment $e_2$ and given post-state $s'$ and finally
\item given environment $e'$ is a result of a push operation on environment $e_2$ updating the value of the split ``SpltModName'' with the one constructed by the computed values $vseq$.
\end{itemize}

The semantics of the split behavior is formalized as follows:

\begin{tabbing}
\textlbrackdbl \=SplMod\textrbrackdbl (e)(e', s, s') $\Leftrightarrow$ 
\\\> $\forall$ \=$e_1, e_2 \in$ Environment, $s_1 \in $ State, vseq $\in$ Value$^*$: \\\>\> \textlbrackdbl SpltParamSeq\textrbrackdbl (e)(e$_1$, s, inState$_\bot$(s$_1$), vseq) $\wedge$
\\\>\>\textlbrackdbl SpltCondSeq\textrbrackdbl (e$_1$) (e$_2$, s$_1$, s') $\wedge$
\\\>\> LET \=s $\in$ Split IN 
\\\>\>\> e' = push(e$_2$, store(inState(s'))(\textlbrackdbl SpltModName\textrbrackdbl(e$_2$)), s(vseq)) 
\\\>\>END
\end{tabbing}

\subsubsection{Split Model Sequence}
The syntactic domain SplModSeq is:\vspace*{0.3cm}

\noindent SplModSeq := EMPTY $|$ (SplMod) SplModSeq

\subsubsection*{Case: EMPTY}
\begin{tabbing}
\textlbrackdbl \=EMPTY\textrbrackdbl (e)(e', s, s') $\Leftrightarrow$ e' = e $\wedge$ s' = inState$_\bot$(s)
\end{tabbing}
\subsubsection*{Case: (SplMod) SplModSeq}
\begin{tabbing}
\textlbrackdbl \=(SplMod) SplModSeq\textrbrackdbl (e)(e', s, s') $\Leftrightarrow$ 
\\\> $\forall$ \= e'' $\in$ Environment, s'' $\in$ State: 
\\\>\>\textlbrackdbl SplMod\textrbrackdbl (e)(e'', s, inState$_\bot$(s'')) $\wedge$ \textlbrackdbl SplModSeq\textrbrackdbl (e'')(e', s'', s')
\end{tabbing}

\subsubsection{Attack Model}
The attack model represents the different types of known/hypothetical attack, their corresponding probabilities and the respective vulnerabilities causing the attack types. The attack model ``AtkModName'' has:
\begin{enumerate}
\item types of attack and their conditional probabilities as specified by the clause \text{:attack-types} and
\item mapping between the types of attack and vulnerabilities as described by the corresponding clause \textbf{:vulnerability-mapping}.
\end{enumerate}
Additionally, the attack model is extended by the rules which map conditional probabilities of the attacks and vulnerabilities. The attack rule ``AtkRulName'' has
\begin{enumerate}
\item a sequence of attack conditions which describe the attack situation as specified by the clause \textbf{if}
\item and the attack consequences which map the probabilities of attacks and vulnerabilities; the maps are
represented by the clause \textbf{then}.
\end{enumerate}
Note that the attack models can be used in the following ways:
\begin{itemize}
\item the models are already known attacks and thus already know the corresponding diagnosis
\item or the models can be hypothetical attacks which can be used to generate rigorous monitors
for the system.
\end{itemize} 

The syntactic domain ``AtkMod'' for the attack model is defined as follows: 
\begin{tabbing}
AtkMod ::= \textbf{define}\=\textbf{-attack-model} AtkModName
\\\> \textbf{:attack-types} (AtkTypeSeq)
\\\> \textbf{:vulnerability-mapping} (AtkVulnrabltyMapSeq)
\end{tabbing}

While the syntactic domain ``AtkRule'' for defining attack rules is defined as follows:
\begin{tabbing}
AtkRule ::= \textbf{defrule} \=AtkRulName (\textbf{:forward})
\\\> \textbf{if} AtkCondSeq
\\\> \textbf{then} AtkConsSeq
\end{tabbing}

Semantically, an attack model results in the environment value ``AtkModel'' as defined below:
\begin{tabbing}
\textlbrackdbl \=AtkMod\textrbrackdbl (e)(e', s, s') $\Leftrightarrow$ 
\\\> $\forall$ \= s'' $\in$ State, aseq, aseq', vnseq $\in$ ISeq, apseq $\in$ Value$^*$: \\\>\> \textlbrackdbl AtkTypeSeq\textrbrackdbl (e)(s, inState$_\bot$(s''), aseq, apseq) $\wedge$ 
\\\>\> \textlbrackdbl AtkVulnrabltyMapSeq\textrbrackdbl (e) (s'', s', aseq', vnseq) $\wedge$ respectsOrder(aseq, aseq') $\wedge$
\\\>\> LET \= amod $\in$ AtkModel IN 
\\\>\>\> e' = push(e\=, store(inState(s'))(\textlbrackdbl AtkModName\textrbrackdbl(e)), amod(aseq, apseq, vnseq)))
\\\>\> END
\end{tabbing}

In detail, the semantics of the syntactic domain ``AtkMod'' updates the environment $e$ with a attack semantic value $amod$ such that
\begin{itemize}
\item in a given environment $e$ and state $s$, the evaluation of ``AtkTypeSeq'' results in a post-state $s''$, a sequence of attack types $aseq$ and a sequence of values (conditional probabilities) $apseq$ and
\item in a given environment $e$ and state $s$, the evaluation of ``AtkVulnrabltyMapSeq'' results in post-state $s'$, a sequence of attack types $aseq'$ and a sequence of vulnerabilities $vnseq$ and
\item the environment $e'$ is an update of environment $e$ with the semantic value $amod$ which is a triple of 
\begin{enumerate}
\item a sequence of attack types,
\item a sequence of corresponding probabilities and
\item a sequence of vulnerabilities causing the attack types, respectively.
\end{enumerate}
\end{itemize}

However, if the semantics of the syntactic domain ``AtkRule'' holds in an environment $e$, then
\begin{itemize}
\item there is some resource $r$ such that (as given in ``AtkCondSeq'' respective ``AtkCond'')
\begin{enumerate}
\item the resource name is ``?resource-name''  and
\item the resource type is ``ResType'' and
\item if the resource has been compromised by an attack ``AtkTypeName'', then
\end{enumerate}
\item the resource $r$ (and its associated component $c$) has behavior as specified by the evaluation of consequences ``AtkCodSeq'' in an environment $e$ and state $s$.
\end{itemize}
Formally, the semantics of the syntactic domain ``AtkRule'' is defined as:
\begin{tabbing}
\textlbrackdbl \=AtkRule\textrbrackdbl (e)(e', s, s') $\Leftrightarrow$ 
\\\> $\exists$ \=r $\in$ Resource, c $\in$ Component: \textlbrackdbl AtkCondSeq\textrbrackdbl (e)(s, s', r, c) $\wedge$
\\\>\> \textlbrackdbl AtkConsSeq\textrbrackdbl (e) (s, s', r, c) $\wedge$ e' = e
\end{tabbing}

\subsubsection{Attack Model Sequence}
The syntactic domain AtkModSeq is:\vspace*{0.3cm}

\noindent AtkModSeq := EMPTY $|$ (AtkMod) AtkModSeq

\subsubsection*{Case: EMPTY}
\begin{tabbing}
\textlbrackdbl \=EMPTY\textrbrackdbl (e)(e', s, s') $\Leftrightarrow$ e' = e $\wedge$ s' = inState$_\bot$(s)
\end{tabbing}
\subsubsection*{Case: (AtkMod) AtkModSeq}
\begin{tabbing}
\textlbrackdbl \=(AtkMod) AtkModSeq\textrbrackdbl (e)(e', s, s') $\Leftrightarrow$ 
\\\> $\forall$ \= e'' $\in$ Environment, s'' $\in$ State: 
\\\>\>\textlbrackdbl AtkMod\textrbrackdbl (e)(e'', s, inState$_\bot$(s'')) $\wedge$ \textlbrackdbl AtkModSeq\textrbrackdbl (e'')(e', s'', s')
\end{tabbing}

\section{Execution Monitor}\label{sec:monitor}
In principle, Architectural Differencer synthesizes both the wrappers and the execution monitor where the wrappers
traces the execution of the target system by creating an event stream (these traces are also called \emph{observations}); while the role of an execution monitor is to interpret the stream against the system (Architectural Model) specification (the execution of the specification is also called \emph{predictions}) by detecting inconsistencies between \emph{observations} and the
\emph{predictions}, if there are any. 

We have already discussed the formal syntax and semantics of the \emph{predictions} in the previous sections, now we first give the formal syntax of the \emph{observations} in this section and the corresponding formal semantics in the following section.

\subsection{Observation Model}
Each runtime event (\emph{observation}) consists of
\begin{itemize}
\item a name ``EvntName'',
\item its type, i.e. \textbf{entry} or \textbf{exit},
\item depending on the type of event
\begin{itemize}
\item either sequence of event parameters (if \textbf{entry} event)
\item or a parameter representing return value of the event (if \textbf{exit} event)
\end{itemize}
\item a numeric value ``Numeral'' representing an operating system level process id, which can be used later to get more information about the event to detect any system level threats and other technical dependencies and
\item a time ``TimeStamp'' of the event which later can be used to detect inconsistencies in the sequence of events.
\end{itemize}

The syntax of the runtime event is defined by the syntactic domain ``Obsrv'' as follow: \vspace*{0.3cm}

\begin{tabbing}
Obsrv := EvntName \= \textbf{entry} $|$ \textbf{exit}
\\\> EvntParamSeq
\\\> Numeral 
\\\> TimeStamp
\end{tabbing}

If the semantics of an observation ``Obsrv'' in a given environment $e$ results in environment $e'$ and transforms a pre-state $s$ into post-state $s'$ then
\begin{itemize}
\item first the parameters are evaluated in a given environment $e$ which results in an environment $e_1$ and sequence of values $pseq$ transforming a given pre-state $s$ into post-state $s_1$ and
\item evaluation of the numeric value ``Numeral'' results in a value $n$ in environment $e_1$ and state $s_1$ and
\item also time stamp ``TimeStamp'' evaluates to a value $t$ in environment $e_1$ and state $s_1$ and finally
\begin{itemize}
\item if the observation is ``entry'' event the resulting environment $e'$ is a result of a push operation on environment $e_2$ updating the value of the observation ``EvntName'' with the semantic  value of the observation, i.e. of type ``RTEvent'' which is constructed with the help of computed input values $pseq$, process id $n$ and time value $t$ and
\item if the observation is ``exit'' event the resulting environment $e'$ is a result of a push operation on environment $e_2$ updating the value of the observation ``EvntName'' with the semantic  value of the observation, i.e. of type ``RTEvent'' which is constructed with the help of computed output values $pseq$ (sequence with a single value), process id $n$ and time value $t$.
\end{itemize}
\end{itemize}

The semantics of the observation is formalized as follows:

\begin{tabbing}
\textlbrackdbl \=Obsrv\textrbrackdbl (e)(e', s, s') $\Leftrightarrow$ 
\\\> $\forall$ \=$e_1, e_2 \in$ Environment, $s_1 \in $ State, pseq $\in$ Value$^*$, n, t $\in$ Value: \\\>\> \textlbrackdbl EvntParamSeq\textrbrackdbl (e)(e$_1$, s, inState$_\bot$(s'), pseq) $\wedge$
\\\>\>\textlbrackdbl Numeral\textrbrackdbl (e$_1$) (inState(s'), n) $\wedge$
\textlbrackdbl TimeStamp\textrbrackdbl (e$_1$) (inState(s'), t) $\wedge$
\\\>\> LET \= rte $\in$ RTEvent IN
\\\>\>\> IF \=isEntry(Obsrv) THEN
\\\>\>\>\> e' = push(e$_2$\=, store(inState(s'))(\textlbrackdbl EvntName\textrbrackdbl(e$_2$))
\\\>\>\>\>\>, rte(pseq, EMPTY, s, s', \textless ``entry'', t, n\textgreater)) 
\\\>\>\> ELSE
\\\>\>\>\> e' = push(e$_2$\=, store(inState(s'))(\textlbrackdbl EvntName\textrbrackdbl(e$_2$))
\\\>\>\>\>\>, rte(EMPTY, pseq, s, s', \textless ``exit'', t, n\textgreater)) 
\\\>\>\>END
\\\>\> END
\end{tabbing}

\subsubsection{Observations}
The event respectively observation stream is a sequence of observations, which is modeled by corresponding syntactic domain ObsrvSeq as follows:\vspace*{0.3cm}

\noindent ObsrvSeq := EMPTY $|$ (Obsrv) ObsrvSeq

The semantics of the observation sequence are similar to the other syntactic sequences discussed earlier in this document.

\subsubsection*{Case: EMPTY}
\begin{tabbing}
\textlbrackdbl \=EMPTY\textrbrackdbl (e)(e', s, s') $\Leftrightarrow$ e' = e $\wedge$ s' = inState$_\bot$(s)
\end{tabbing}
\subsubsection*{Case: (Obsrv) ObsrvSeq}
\begin{tabbing}
\textlbrackdbl \=(Obsrv) ObsrvSeq\textrbrackdbl (e)(e', s, s') $\Leftrightarrow$ 
\\\> $\forall$ \= e'' $\in$ Environment, s'' $\in$ State: 
\\\>\>\textlbrackdbl Obsrv\textrbrackdbl (e)(e'', s, inState$_\bot$(s'')) $\wedge$ \textlbrackdbl ObsrvSeq\textrbrackdbl (e'')(e', s'', s')
\end{tabbing}

\section{Semantics of the Execution Monitor}\label{sec:monitor-semantics}
Though the technical details of the operation of the execution monitor are discussed in~\cite{Shrobe:2006}, in the following we give their informal semantics.

We presume that a reasonable fine grained level behavior of the target system is specified in the corresponding
System Architectural Model. When the target system starts execution, an initial ``startup'' event is generated and
dispatched to the top level component (module) of the system which transforms the execution state of the component
into ``running'' mode. The component instantiates its subnetwork (of components, if there is one) and also propagates
the data along its data links by enabling the corresponding control links (if involved). When the data arrives on the input port of the component, the execution monitor checks if it is complete; if so, the execution monitor checks the preconditions of the component for the data and if they succeed, it transform the state of the component into ``ready'' mode. In case, any of the preconditions fails, it enables diagnosis engine.

After the above startup of the target system, the execution monitor starts monitoring the arrival of every \emph{observation} (runtime event) as follows:
\begin{enumerate}
\item If the event is a ``method entry'', then the execution monitor checks if this is one of the ``entry events'' of the
corresponding component in the ``ready'' state; if so, then after receiving the data and the respective preconditions are checked; if they succeed, then the data is applied on the input port of the component and the mode of the execution state is changed to ``running''.
\item If the event is a ``method exit'', then the execution monitor checks if this one of the ``exit events'' of the component in the ``running'' state; if so, it changes its state into ``completed'' mode and collects the data from the output port of the component and checks for the corresponding postconditions. Should the checks fail, the execution monitor enables the diagnosis engine.
\item If the event is one of the ``allowable events'' of the component, it continues execution and finally
\item if the event is an unexpected event, i.e. it is neither an ``entry event'', nor an ``exit event'' and also not in ``allowable events'', then the execution monitor starts diagnosis.
\end{enumerate}

Based on the above behavioral description of the execution monitor, we have formalized the corresponding semantics of the execution monitor as follows:

\begin{tabbing}
$\forall$ \= app $\in$ Target\_System, sam $\in$ System\_Architectural\_Model, c $\in$ Component,
\\\> s, s' $\in$ State, t, t' $\in$ State$_s$, d, d' $\in$ Environment$_s$, e, e' $\in$ Environment, rte $\in$ RTEvent:
\\\> \textlbrackdbl sam\textrbrackdbl(d)(d', t, t') $\wedge$ \textlbrackdbl app\textrbrackdbl(e)(e', s, s') $\wedge$ startup(s, app) $\wedge$ isTop(c, \textlbrackdbl app\textrbrackdbl(e)(e', s, s')) $\wedge$
\\\> setMode(s, ``running'') $\wedge$ arrives(rte, s)  $\wedge$ equals(t, s) $\wedge$ equals(d, e)
\\\> $\Rightarrow$ \=
\\\>\> $\forall$ \= p, p' $\in$ Environment$^*$, m, n $\in$ State$_\bot^*$: 
\\\>\>\> equals(m(0), s) $\wedge$ equals(p(0), e)
\\\>\>\> $\Rightarrow$ \=
\\\>\>\>\> $\exists$ \= k $\in$ $\mathbb{N}$, p, p' $\in$ Environment$^*$, m, n $\in$ State$_\bot^*$:  
\\\>\>\>\>\> $\forall$ \=i $\in$ $\mathbb{N}_k$ : monitors(i, rte, c, p, p', m, n) $\wedge$
\\\>\>\>\>\>\> ( eq\=Mode(n(k), ``completed'') $\wedge$ eqFlag(n(k), ``normal'')  $\wedge$ 
\\\>\>\>\>\>\>\> equals(s', n(k))
\\\>\>\>\>\>\> $\vee$
\\\>\>\>\>\>\> eq\=Flag(n(k), ``compromised'') 
\\\>\>\>\>\>\>\> $\Rightarrow$ \=
\\\>\>\>\>\>\>\>\> enableDiagnosis(p'(k))(n(k), inBool(\texttt{true}))  $\wedge$ equals(s', n(k)) )
\end{tabbing}


In detail, given a target system ``app'' and its specification``sam'' and their semantices are defined such that their corresponding pre-states are equivalent. Furthermore, if the application starts ``startup(...)'', and an arbitrary $c$ is a top-level component ``isTop(...)'', then the current state of the component is marked as ``running'' and when an
observation ``rte'' arrives in this state, then the monitor starts monitoring the event stream/sequence and thus, here, we have formalized the corresponding semantics of the monitor by the two sequences of
pre- and post-states~\cite{Khan:2012c} and their respective sequences of the pre- and post-environments. Both the former and later sequences are constructed from their corresponding pre- and post objects. The arrival and monitoring of the $ith$ observation (event) transforms state $pre(i)$ into state $post(i+1)$ from which the state $pre(i+1)$ is constructed and the same repeats for the construction of the corresponding environments. No event can be accepted in an $Error$ state and the corresponding monitoring terminates either when the application has terminated with ``normal'' mode or when there is some misbehavior is detected as indicated by the respective ``compromised'' state. This semantics is formalized with the help of predicate ``monitor'', for details please see Subsection~\ref{subsec:afp}.

Finally, when there are sequences of states and environments for which the predicate ``monitor'' holds, then either the
given post-state $s'$ is equal to the ``monitor''ed post-state ``n(k)'' which is in ``completed'' mode and has a ``normal'' flag or post-state ``n(k)''  is ``compromised'' and in this case diagnosis is enabled which successfully transforms the compromised state into a normal state which results in the given post-state $s'$.

%

\section{Conclusions and Future Work}\label{sec:conclusions}
In this report, we gave the formal definition of the syntax and semantices of the System Architectural Model and
the Execution Monitor of AWDRAT. These definitions help to understand internal behavior of the corresponding components on one hand, and also serves as a formal basis for ADWRAT to extend the current system on the other hand. Based on this formalism, we are currently working on the formal reliability (soundness) analysis of the Execution Monitor of AWDRAT.

In future, we plan to extend AWDRAT such that a target system behavior is specified using Abstract State Machine (ASM)~\cite{ASM:2003} based formalism which then will automatically translate into a semantically equivalent System Architectural Model. This will allow to already check the inconsistencies in the system behavior with existing ASM supported tools~\cite{DKAL:2013}.

\section*{Acknowledgment}
The authors cordially thank Adam Chilpala for his  valuable and constructive remarks and suggestions.

\newpage
\bibliography{sam}
\bibliographystyle{plain}

\newpage
\appendix
\appendixpage
Appendix A gives the formal abstract syntax (language grammar) for the specification language ``system architectural model'' of AWDRAT.
\addappheadtotoc
\section{Formal Syntax of System Architectural Model}\label{sec:sam-syntax}
\subsection{Declaration of Syntactic Domains}
/* top level syntactic domains */\\
SAM $\in$ System\_Architectural\_Model \\
RegModSeq $\in$ Register\_Model\_Sequence\\
StrModSeq $\in$ Structural\_Model\_Sequence\\
BehModSeq $\in$ Behavioral\_Model\_Sequence\\
SplModSeq $\in$ Split\_Model\_Sequence\\
AtkModSeq $\in$ Attack\_Model\_Sequence\\
AtkRuleSeq $\in$ Attack\_Rule\_Sequence\\[3mm]
/* top level syntactic sub-domains */\\
RegMod $\in$ Register\_Model\\
StrMod $\in$ Structural\_Model\\
BehMod $\in$ Behavioral\_Model\\
SplMod $\in$ Split\_Model\\
AtkMod $\in$ Attack\_Model\\
AtkRule $\in$ Attack\_Rule\\[3mm]
/* event related syntactic domains */\\
Evnt $\in$ Event\\
EvntSeq $\in$ Event\_Sequence\\
EvntName $\in$ Event\_Name\\
EvntNameSeq $\in$ Event\_Name\_Sequence\\
EvntParamSeq $\in$ Event\_Parameter\_Sequence\\[3mm]
/* java related syntactic domains */\\
JavClaName $\in$ Java\_Class\_Name\\
JavMetName $\in$ Java\_Method\_Name\\
JavParam $\in$ Java\_Parameter\\
JavParamSeq $\in$ Java\_Parameter\_Sequence\\
JavParamName $\in$ Java\_Parameter\_Name\\
JavParamType $\in$ Java\_Parameter\_Type\\[3mm]
/* object related syntactic domains */\\
ObjName $\in$ Object\_Name\\
ObjtNameStr $\in$ Object\_Name\_String\\
ObjNameSeq $\in$ Object\_Name\_Sequence\\
ObjType $\in$ Object\_Type\\
ObjComp $\in$ Object\_Component\\
ObjCompSeq $\in$ Object\_Component\_Sequence\\[3mm]
/* behavioral condition, parameter and situation related syntactic domains */\\
BehCond $\in$ Behavioral\_Condition\\
BehCondSeq $\in$ Behavioral\_Condition\_Sequence\\
BehCondMode $\in$ Behavioral\_Condition\_Mode\\
BehParam $\in$ Behavioral\_Parameter\\
BehParamSeq $\in$ Behavioral\_Parameter\_Sequence\\
BehSit $\in$ Behavioral\_Situation\\[3mm]
/* branch related syntactic domains */\\
BrnchName $\in$ Branch\_Name\\
BrnchNameSeq $\in$ Branch\_Name\_Sequence\\
BrnchCond $\in$ Branch\_Condition\\[3mm]
/* component related syntactic domains */\\
Comp $\in$ Component\\
CompSeq $\in$ Component\_Sequence\\
CompName $\in$ Component\_Name\\
CompType $\in$ Component\_Type\\[3mm]
/* control flow related syntactic domains */\\
CtrlFlow $\in$ Control\_Flow\\
CtrlFlowSeq $\in$ Control\_Flow\_Sequence\\[3mm]
/* function related syntactic domains */\\
FuncName $\in$ Function\_Name\\
FuncParam $\in$ Function\_Parameter\\
FuncParamSeq $\in$ Function\_Parameter\_Sequence\\[3mm]
/* split related syntactic domains */\\
SpltCF $\in$ Split\\
SpltCFSeq $\in$ Split\_Sequence\\
SpltName $\in$ Split\_Name\\
SpltModName $\in$ Split\_Model\_Name\\
SpltParamSeq $\in$ Split\_Parameter\_Sequence\\
SpltCond $\in$ Split\_Condition\\
SpltCondSeq $\in$ Split\_Condition\_Sequence\\[3mm]
/* join related syntactic domains */\\
JoinCF $\in$ Join\\
JoinCFSeq $\in$ Join\_Sequence\\
JoinName $\in$ Join\_Name\\
JoinParamSeq $\in$ Join\_Parameter\_Sequence\\[3mm]
/* data flow related syntactic domains */\\
DataFlow $\in$ Data\_Flow\\
DataFlowSeq $\in$ Data\_Flow\_Sequence\\[3mm]
/* resource related syntactic domains */\\
Res $\in$ Resource\\
ResSeq $\in$ Resource\_Sequence\\
ResName $\in$ Resource\_Name\\
ResType $\in$ Resource\_Type\\
ResMap $\in$ Resource\_Mapping\\
ResMapSeq $\in$ Resource\_Mapping\_Sequence\\
ResModMap $\in$ Resource\_Model\_Mapping\\[3mm]
/* model mapping syntactic domains */\\
ModMap $\in$ Model\_Mapping\\
ModMapSeq $\in$ Model\_Mapping\_Sequence\\[3mm]
/* vulnerability related syntactic domains */\\
Vulnrablty $\in$ Vulnerability\\
VulnrabltyName $\in$ Vulnerability\_Name\\
VulnrabltySeq $\in$ Vulnerability\_Sequence\\[3mm]
/* attack related syntactic domains */\\
AtkType $\in$ Attack\_Type\\
AtkTypeSeq $\in$ Attack\_Type\_Sequence\\
AtkModName $\in$ Attack\_Model\_Name\\
AtkCond $\in$ Attack\_Condition\\
AtkCondSeq $\in$ Attack\_Condition\_Sequence\\
AtkCons $\in$ Attack\_Consequence\\
AtkConsSeq $\in$ Attack\_Consequence\_Sequence\\
AtkTypeName $\in$ Attack\_Type\_Name\\
AtkRulName $\in$ Attack\_Rule\_Name\\
AtkVulnrabltyMap $\in$ Attack\_Vulnerability\_Mapping\\
AtkVulnrabltyMapSeq $\in$ Attack\_Vulnerability\_Mapping\_Sequence\\[3mm]
/* other syntactic domains */\\
MembName $\in$ Member\_Name\\
ParamName $\in$ Parameter\_Name\\
DSCond $\in$ Data\_Structure\_Condition\\
ISeq $\in$ Identifier\_Sequence\\

\subsection{Grammar}
Based on the declarations of various syntactic domains, in this section we discuss the grammar rules
for the domains.
\begin{tabbing}
/* top level syntactic domains */\\
SAM ::= RegModSeq StrModSeq BehModSeq SplModSeq\\
RegModSeq ::= \=EMPTY 
\\\>$|$ (RegMod) RegModSeq\\
StrModSeq ::= \=EMPTY 
\\\>$|$ (StrMod) StrModSeq\\
BehModSeq ::= \=EMPTY 
\\\>$|$ (BehMod) BehModSeq\\
SplModSeq ::= \=EMPTY 
\\\>$|$ (SplMod) SplModSeq\\
AtkModSeq ::= \=EMPTY 
\\\>$|$ (AtkMod) AtkModSeq\\
AtkRuleSeq ::= \=EMPTY 
\\\>$|$ (AtkRule) AtkRuleSeq\\[3mm]
/* top level syntactic sub-domains */\\
RegMod ::= \textbf{register-event} \='EvntName JavClaName JavMetName '(JavParamSeq)
\\\>\=[ \textbf{:static} ObjName ] 
\\\>\>[ \textbf{:output-type} JavParam ] 
\\\>\>[ \textbf{:bypass} ObjNameStr ]
\\\>\>[ \textbf{:}EvntName ObjName]\\
StrMod ::= \textbf{define-ensemble} \=CompName
\\\> \textbf{:entry-events} \textbf{:auto} $|$ (EvntSeq)
\\\> \textbf{:exit-events} (EvntSeq)
\\\> \textbf{:allowable-events} (EvntSeq)
\\\> \textbf{:inputs} (ObjNameSeq)
\\\> \textbf{:outputs} (ObjNameSeq)
\\\> \textbf{:components} (CompSeq)
\\\> \textbf{:controlflows} (CtrlFlowSeq)
\\\> \textbf{:splits} (SpltCFSeq)
\\\> \textbf{:joins} (JoinCFSeq)
\\\> \textbf{:dataflows} (DataFlowSeq)
\\\> \textbf{:resources} (ResSeq)
\\\> \textbf{:resource-mapping} (ResMapSeq)
\\\> \textbf{:model-mappings} (ModMapSeq)
\\\> \textbf{:vulnerabilities} (VulnrabltySeq)\\
BehMod ::= \textbf{defbehavior-model} \=(CompName \textbf{normal} $|$ \textbf{compromised})
\\\> \textbf{:inputs} (ObjNameSeq)
\\\> \textbf{:outputs} (ObjNameSeq)
\\\> \textbf{:allowable-events} (EvntSeq)
\\\> \textbf{:prerequisites} (BehCondSeq)
\\\> \textbf{:postconditions} (BehCondSeq)\\
SplMod ::= \textbf{defsplit} SpltModName\textbf{?} (SpltParamSeq) SpltCondSeq)\\
AtkMod ::= \textbf{define-attack-model} \=AtkModName
\\\> \textbf{:attack-types} (AtkTypeSeq)
\\\> \textbf{:vulnerability-mapping} (AtkVulnrabltyMapSeq)\\
AtkRule ::= \textbf{defrule} \=AtkRulName (\textbf{:forward})
\\\> \textbf{if} AtkCondSeq
\\\> \textbf{then} AtkConsSeq\\[3mm]
/* event related syntactic domains */\\
Evnt ::= EvntName $|$ (EvntName [\textbf{entry $|$ exit}] (EvntParamSeq))\\
EvntSeq ::= \=EMPTY 
\\\>$|$ Evnt EvntSeq\\
EvntParam ::= \=I Iseq \\\>$|$ \textbf{nil} I Iseq \\\>$|$ I ISeq \textbf{nil}\\
EvntParamSeq ::= \=EMPTY 
\\\>$|$ EvntParam EvntParamSeq\\[3mm]
/* java related syntactic domains */\\
JavClaName ::= "I"\\
JavMetName ::= "ID" $|$ ''\textless I\textgreater``\\
JavParam ::= (JavParamType JavParamName)\\
JavParamSeq ::= \=EMPTY \\\>$|$ JavParam JavaParamSeq\\
JavParamName ::= ''I``\\
JavParamType ::= ''ID`` $|$ "ID[]"\\[3mm]
/* object related syntactic domains */\\
ObjComp ::= (ObjName CompName\textbf{?})\\
ObjCompSeq ::= \=EMPTY \\\>$|$ ObjComp ObjCompSeq\\
ObjNameStr ::= ObjName $|$ ''ObjName``\\[3mm]
/* behavioral condition related syntactic domains */\\
BehCond ::= \=[ DSCond ObjName ObjType BehCondMode ]
\\\> $|$ [ \textbf{and} BehCond ]
\\\> $|$ [ \textbf{or} BehCond ]
\\\> $|$ [ \textbf{not} BehCond ]
\\\> $|$ [ SpecFuncName BehParamSeq BehSit ]\\
BehCondSeq ::= \=EMPTY \\\>$|$ BehCond BehCondSeq\\
BehCondMode ::= EMPTY $|$ \textbf{good}\\
BehParam ::= \textbf{?}ObjName $|$ (MembName \textbf{?}ObjName)\\
BehParamSeq ::= \=EMPTY \\\>$|$ BehParam BehParamSeq\\
BehSit ::= \textbf{?before-}CompName $|$ \textbf{?after-}CompName\\[3mm]
/* branch condition syntactic domain */\\
BrnchCond ::= FuncName FuncParamSeq\\[3mm]
/* component related syntactic domains */\\
Comp ::= (CompName \textbf{:type} CompType \textbf{:models} (\textbf{normal} [\textbf{compromised}]))\\
CompSeq ::= \=EMPTY 
\\\>$|$ Comp CompSeq\\[3mm]
/* control flow related syntactic domains */\\
CtrlFlow ::= (\textbf{before} $|$ \textbf{after} CompName[\textbf{?-}BrnchName])\\
CtrlFlowSeq ::= \=EMPTY \\\>$|$ CtrlFlow CtrlFlowSeq\\[3mm]
/* function related syntactic domains */\\
FuncParam ::= \textbf{?}ParamName $|$ \textbf{'}ParamName $|$ ParamName\textbf{?} $|$ \textbf{not} (ParamName)\\
FuncParamSeq ::= \=EMPTY \\\>$|$ FuncParam FuncParamSeq\\[3mm]
/* split of control flow related syntactic domains */\\
SpltCF ::= (SpltName\textbf{?} SpltModName\textbf{?} [(SpltParamSeq)] (BrnchNameSeq))\\
SpltCFSeq ::= \=EMPTY \\\>$|$ SpltCF SpltCFSeq\\
SpltCondSeq ::= \=EMPTY \\\>$|$ SpltCond SplitCondSeq\\
SpltCond ::= (BrnchName (BrnchCond))\\\\[3mm]
/* join of control flow related syntactic domains */\\
JoinCF ::= (JoinName\textbf{?} [(JoinParamSeq)] (BrnchNameSeq))\\
JoinCFSeq ::= \=EMPTY \\\>$|$ JoinCF JoinCFSeq\\[3mm]
/* data flow related syntactic domains */\\
DataFlow ::= (ObjCompSeq)\\
DataFlowSeq ::= \=EMPTY \\\>$|$ DataFlow DataFlowSeq\\[3mm]
/* resource related syntactic domains */\\
Res ::= (ResName ResType [(\textbf{normal} $|$ \textbf{hacked} FVal)]+)\\
ResSeq ::= \=EMPTY \\\>$|$ Res ResSeq\\
ResType ::= File $|$ Port $|$ Mem\\
ResMap ::= (CompName ResName)\\
ResMapSeq ::= \=EMPTY \\\>$|$ ResMap ResMapSeq\\
ResModMap ::= \=ResName \textbf{normal} $|$ \textbf{hacked} FVal 
\\\>$|$ ((ResName \textbf{normal} $|$ \textbf{hacked})) FVal\\[3mm]
/* model mapping related syntactic domains */\\
ModMap ::= (CompName \textbf{noromal} $|$ \textbf{compromised} ResModMap)\\
ModMapSeq ::= \=EMPTY \\\>$|$ ModMap ModMapSeq\\[3mm]
/* vulnerability related syntactic domains */\\
Vulnrablty ::= (ResName VulnrabltyName)\\
VulnrabltySeq ::= \=EMPTY \\\>$|$ Vulnrablty VulnrabltySeq\\[3mm]
/* attack related syntactic domains */\\
AtkType ::= (AtkTypeName FVal)\\
AtkTypeSeq ::= \=EMPTY \\\>$|$ AtkType AtkTypeSeq\\
AtkCond ::= \=[\textbf{resource} \textbf{?}ensemble \textbf{?}ResName \textbf{?}Res]
\\\> [\textbf{resource-type-of} \textbf{?}Res ResType]
\\\> [\textbf{resource-might-have-been-attacked} \textbf{?}Res AtkTypeName]
\\\> $|$ [ \textbf{and} AtkCond ]
\\\> $|$ [ \textbf{or} AtkCond ]
\\\> $|$ [ \textbf{not} AtkCond ]\\
AtkCondSeq ::= \=EMPTY \\\>$|$ AtkCond AtkCondSeq\\
AtkCons ::= \=[\textbf{attack-implies-compromised-mode} \=AtkTypeName \textbf{?}Res 
\\\>\> \textbf{normal} $|$ \textbf{compromised} FVal]
\\\> $|$ [ \textbf{and} AtkCons ]
\\\> $|$ [ \textbf{or} AtkCons ]
\\\> $|$ [ \textbf{not} AtkCons ]\\
AtkConsSeq ::= \=EMPTY \\\>$|$ AtkCons AtkConsSeq\\
AtkVulnrabltyMap ::= (VulnrabltyName AtkTypeName)\\
AtkVulnrabltyMapSeq ::= \=EMPTY \\\>$|$ AtkVulnrabltyMap AtkVulnrabltyMapSeq\\[3mm]
/* syntactic domains of various names and types */\\
CompName, \=CompType, FuncName, ObjName, ObjType, 
\\\>EvntName, SpltName, SpltModName ::= I\\
JoinName, \=ResName, BrnchName, VulnrabltyName, SpecFuncName
\\\>, ParamName, MembName ::= I\\
AtkModName, \=AtkTypeName, AtkRulName ::= I\\[3mm]
/* syntactic domains of various sequences */\\
ObjNameSeq, SpltParamSeq, JoinParamSeq, BrnchNameSeq ::= ISeq\\[3mm]
/* other syntactic domains */\\
DSCond ::= EMPTY $|$ \textbf{dscs}\\
ISeq :: = \=EMPTY \\\>$|$ I ISeq \\
I ::= any valid LISP system name \\
FVal ::= a sequence of decimal digits prefixed by a period (valid float value)
\end{tabbing}


\section{An Example of a System Architectural Model}\label{sec:example}
In this section, we give the syntax of an example System Architectural Model of MAF editor system which is discussed in detail in~\cite{Shrobe:2006}. In the following, we give a brief detail on how to read the example, i.e. \verb|maf-editor| is the top level component of the application whose structural behavior is specified at first. Every sentence of the specification is self-explanatory. In principle, the behavior of every component in the subnetwork of a parent component has to be specified separately with two corresponding parts, e.g. a component \verb|maf-startup| (which is in the subnetwork of top-level component as mentioned in \verb|:components| clause) has
\begin{enumerate}
\item structural behavior as specified by clause \begin{center}\verb|define-ensemble maf-startup| \end{center}
\item normal behavior as specified by the clause \begin{center}\verb|defbehavior-model (maf-startup norml)|\end{center}
\item and a corresponding compromised behavior is specified by the clause \begin{center}\verb|defbehavior-model (maf-startup compromised)|\end{center}
\end{enumerate}

The former part corresponds to the \emph{control} level of the specification while the latter two corresponds to the \emph{behavioral} level of the specification of the component.

Furthermore, as explained in Section~\ref{sec:sam}, the split behavior of the component \verb|maf-create-events| is further specified with the corresponding clauses, e.g. \begin{center}
\verb|defsplit maf-more-events?|
\end{center}
Also, any pre/postcondition that is followed by the \verb|dscs| specifies the data structure consistency property.

\subsection{MAF Editor Model}
\begin{verbatim}
(define-ensemble maf-editor
    :entry-events :auto
    :inputs ()
    :outputs (the-model)
    :components ((startup :type maf-startup :models (normal compromised))
    				(create-model :type maf-create-model :models (normal compromised))
    				(create-events :type maf-create-events :models (normal compromised))
    				(save :type maf-save :models (normal compromised)))
    				
    :controlflows ((before maf-editor before startup)
    				(after startup before create-model))
    				
    :dataflows ((the-model create-model the-model create-events)
    			(the-model create-events the-model save)
    			(the-model save the-model maf-save-model))
    			
    :resources ((imagery image-file (normal .7) (hacked .3)) 
    			(code-files loadable-files (normal .8) (hacked .2)))
    			
    :resource-mappings ((startup imagery)
    					(create-model code-files)
    					(create-events code-files)
    					(save-model code-files))
    					

    :model-mappings ((startup normal ((imagery normal)) .99)
		     (startup compromised ((imagery normal)) .01)
		     (startup normal ((imagery hacked)) .9)
		     (startup compromised ((imagery hacked)) .1)
		     
		     (create-model normal ((code-files normal)) .99)
		     (create-model compromised ((code-files normal)) .01)
		     (create-model normal ((code-files hacked)) .9)
		     (create-model compromised ((code-files hacked)) .1)
		     
		     (create-events normal ((code-files normal)) .99)
		     (create-events compromised ((code-files normal)) .01)
		     (create-events  normal ((code-files hacked)) .9)
		     (create-events compromised ((code-files hacked)) .1)
		     
		     (save normal ((code-files normal)) .99)
		     (save compromised ((code-files normal)) .001)
		     (save normal ((code-files hacked)) .01)
		     (save compromised ((code-files hacked)) .999))
    
    :vulnerabilities ((imagery reads-complex-imagery)
		      (code-files loads-code)
		      ))
		 
(define-ensemble maf-startup
    :entry-events (startup)
    :exit-events (startup)
    :allowable-events (post-validate create-client-frame 
				     center-action load-image)
    :inputs ()
    :outputs ())

(defbehavior-model (maf-startup normal)
    :inputs ()
    :outputs ()
    :prerequisites ()
    :post-conditions ())

(defbehavior-model (maf-startup compromised)
    :inputs ()
    :outputs ()
    :prerequisites ()
    :post-conditions ())

;;; Need defbehaviors for each of these even if its empty

(define-ensemble maf-create-model
    :entry-events (create-mission-action-action-performed)
    :exit-events (mission-builder-submit)
    :allowable-events (create-mission-builder-with-client-panel
		       create-mission-builder
		       create-mission-builder-with-hash-table
		       mission-builder-submit
		       (set-initial-info exit (the-model nil))
		       create-mission-action-action-performed
		       retrieve-info
		       create-mission-action-action-performed
		       (set-initial-info entry)
		       )
    :inputs ()
    :outputs (the-model))

(defbehavior-model (maf-create-model normal)
    :inputs ()
    :outputs (the-model)
    :prerequisites ()
    :post-conditions ([dscs ?the-model mission-builder good])
    )

(defbehavior-model (maf-create-model compromised)
    :inputs ()
    :outputs (the-model)
    :prerequisites ()
    :post-conditions ([not [dscs ?the-model mission-builder good]])
    )

(define-ensemble maf-create-events
    :entry-events :auto 
    :exit-events ()
    :allowable-events ()
    :inputs (the-model)
    :outputs (the-model)
    :components ((get-next-cmd :type maf-get-next-cmd :models (normal))
		 (get-event-info :type maf-get-event-info :models (normal compromised))
		 (add-event-to-model :type maf-add-event-to-model :models 
		 					(normal compromised))
		 (get-leg :type maf-get-leg :models (normal compromised))
		 (get-movement :type maf-get-movement :models (normal compromised))
		 (get-sortie :type maf-get-sortie :models (normal compromised))
		 (add-additional-info-to-model :type maf-add-additional-info :models 
		 						(normal compromised))
		 (continue :type maf-create-events :models (normal compromised)))
		 
    :dataflows ((the-model maf-create-events the-model join-exit-exit)
		(the-model maf-create-events the-model add-event-to-model)
		(the-cmd get-next-cmd cmd more-events?)
		(the-event get-event-info the-event add-event-to-model)
		(the-model add-event-to-model the-model join-events-non-take-off)
		(the-event get-event-info event takeoff?)
		(the-leg get-leg the-leg add-additional-info-to-model)
		(lms-event-counter get-leg event-number add-additional-info-to-model)
		(the-movement get-movement the-movement add-additional-info-to-model)
		(the-sortie get-sortie the-sortie add-additional-info-to-model)
		(the-model add-event-to-model the-model add-additional-info-to-model)
		(the-model add-additional-info-to-model the-model join-events-take-off)
		(the-model join-events the-model continue)
		(the-model continue the-model join-exit-recur)
		(the-model join-exit the-model maf-create-events)
		)
		
   :controlflows ((after more-events?-build-event before add-event-to-model)
   				(after more-events?-exit before join-exit-exit)
   				(after takeoff?-get-additional-info before get-leg)
   				(after takeoff?-get-additional-info before get-movement)
   				(after takeoff?-get-additional-info before get-sortie)
   				(after takeoff?-exit before join-events-non-take-off))
		   
    :splits ((more-events? maf-more-events? (cmd) (build-event exit))
	     (takeoff? maf-takeoff? (event) (get-additional-info exit)))
	     
    :joins ((join-events (the-model) (take-off non-take-off))
	    (join-exit (the-model) (recur exit)))

    :resources ((code-files loadable-files (normal .8) (hacked .2)))

    :resource-mappings ((get-event-info  code-files)
    						(add-event-to-model code-files)
    						(get-leg code-files)
    						(get-movement code-files)
    						(get-sortie code-files)
    						(add-additional-info-to-model code-files)
    						(continue code-files))

    :model-mappings ((get-event-info normal code-files normal .99)
    					(get-event-info compromised code-files normal .01)
    					(get-event-info normal code-files hacked .9)
    					(get-event-info compromised code-files hacked .1)
		     
		     (add-event-to-model normal code-files normal .99)
		     (add-event-to-model compromised code-files normal .01)
		     (add-event-to-model  normal code-files hacked .9)
		     (add-event-to-model compromised code-files hacked .1)
		     
		     (get-leg normal code-files normal .99)
		     (get-leg compromised code-files normal .001)
		     (get-leg normal code-files hacked .01)
		     (get-leg compromised code-files hacked .999)
		     
		     (get-movement normal code-files normal .99)
		     (get-movement compromised code-files normal .001)
		     (get-movement normal code-files hacked .01)
		     (get-movement compromised code-files hacked .999)
		     
		     (get-sortie normal code-files normal .99)
		     (get-sortie compromised code-files normal .001)
		     (get-sortie normal code-files hacked .01)
		     (get-sortie compromised code-files hacked .999)
		     
		     (add-additional-info-to-model normal code-files normal .99)
		     (add-additional-info-to-model compromised code-files normal .001)
		     (add-additional-info-to-model normal code-files hacked .01)
		     (add-additional-info-to-model compromised code-files hacked .999)
		     
		     (continue normal code-files normal .99)
		     (continue compromised code-files normal .001)
		     (continue normal code-files hacked .01)
		     (continue compromised code-files hacked .999))
		     
    :vulnerabilities ((code-files loads-code))
    )

(defbehavior-model (maf-create-events normal)
    :inputs (the-model)
    :outputs (the-model)
    :prerequisites ([dscs ?the-model mission-builder good])
    :post-conditions ([dscs ?the-model mission-builder good])
    )

(defbehavior-model (maf-create-events compromised)
    :inputs (the-model)
    :outputs (the-model)
    :prerequisites ([dscs ?the-model mission-builder good])
    :post-conditions ([not [dscs ?the-model mission-builder good]])
    )

(define-ensemble maf-get-next-cmd
    :entry-events (next-cmd)
    :exit-events ((next-cmd exit (the-cmd)))
    :inputs ()
    :outputs (the-cmd))

(defbehavior-model (maf-get-next-cmd normal)
    :inputs ()
    :outputs (the-cmd)
    :prerequisites ()
    :post-conditions ())

(define-ensemble maf-get-event-info
    :entry-events (create-mission-event-point)
    :allowable-events (set-current-point 
		       (create-mission-event-point exit)
		       create-mission-event-object
		       meo-set-information 
		       mpl-action-performed
		       close-form 
		       add-new-event-internal)
    :exit-events ((got-event-info exit (the-event)))
    :inputs ()
    :outputs (the-event))

(defbehavior-model (maf-get-event-info normal)
    :inputs ()
    :outputs (the-event)
    :prerequisites ()
    :post-conditions ([dscs ?the-event event good]))

(defbehavior-model (maf-get-event-info compromised)
    :inputs ()
    :outputs (the-event)
    :prerequisites ()
    :post-conditions ([not [dscs ?the-event event good]]))

(define-ensemble maf-add-event-to-model
    :entry-events (update-msn-evt)
    :allowable-events 
    ((update-msn-evt exit (mb event-number event))
     add-new-event-internal
     create-new-additional-mission-info-panel
     )
    :exit-events (mpl-action-performed)
    :inputs (the-event the-model)
    :outputs (the-model event-number))

(defbehavior-model (maf-add-event-to-model normal)
    :inputs (the-event the-model)
    :outputs (the-model event-number)
    :prerequisites ([dscs ?the-event event good]
		    [dscs ?the-model mission-builder good])
    :post-conditions 
    ([add-to-map (events ?the-model)?event-number ?the-event
		 ?before-maf-add-event-to-model]
     [dscs ?the-model mission-builder good]))

(defbehavior-model (maf-add-event-to-model compromised)
    :inputs (the-event the-model)
    :outputs (the-model event-number)
    :prerequisites ([not [dscs ?the-event event good]]
		    [not [dscs ?the-model mission-builder good]])
    :post-conditions 
    ([dscs ?the-model mission-builder good]))

(define-ensemble maf-get-leg
    :entry-events (retrieve-leg)
    :exit-events ((retrieve-leg exit (nil the-leg lms-event-counter)))
    :allowable-events (create-mission-leg-object mlo-set-information)
    :inputs ()
    :outputs (the-leg lms-event-counter))

(defbehavior-model (maf-get-leg normal)
    :inputs ()
    :outputs (the-leg lms-event-counter)
    :prerequisites ()
    :post-conditions ([dscs ?the-leg leg good]))

(defbehavior-model (maf-get-leg compromised)
    :inputs ()
    :outputs (the-leg lms-event-counter)
    :prerequisites ()
    :post-conditions ([not [dscs ?the-leg leg good]]))

(define-ensemble maf-get-movement
    :entry-events (retrieve-movement)
    :exit-events ((retrieve-movement exit (nil the-movement)))
    :allowable-events 
    (create-mission-movement-object mmo-set-information)
    :inputs ()
    :outputs (the-movement))

(defbehavior-model (maf-get-movement normal)
    :inputs ()
    :outputs (the-movement)
    :prerequisites ()
    :post-conditions ([dscs ?the-movement movement good]))

(defbehavior-model (maf-get-movement compromised)
    :inputs ()
    :outputs (the-movement)
    :prerequisites ()
    :post-conditions ([not [dscs ?the-movement movement good]]))

(define-ensemble maf-get-sortie
    :entry-events (retrieve-sortie)
    :exit-events ((retrieve-sortie exit (nil the-sortie)))
    :allowable-events 
    (create-mission-sortie-object mso-set-information)
    :inputs ()
    :outputs (the-sortie))

(defbehavior-model (maf-get-sortie normal)
    :inputs ()
    :outputs (the-sortie)
    :prerequisites ()
    :post-conditions ([dscs ?the-sortie sortie good]))

(defbehavior-model (maf-get-sortie compromised)
    :inputs ()
    :outputs (the-sortie)
    :prerequisites ()
    :post-conditions ([not [dscs ?the-sortie sortie good]]))

(define-ensemble maf-add-additional-info
    :entry-events ((retrieve-sortie exit))
    :exit-events (Mission-builder-add-info)
    :inputs (the-model the-leg the-movement the-sortie event-number)
    :outputs (the-model))

(defbehavior-model (maf-add-additional-info normal)
    :inputs (the-model the-leg the-movement the-sortie event-number)
    :outputs (the-model)
    :prerequisites  ([dscs ?the-leg leg good]
		     [dscs ?the-movement movement good]
		     [dscs ?the-sortie sortie good]
		     [dscs ?the-model mission-builder good])
    :post-conditions ([add-to-map (legs ?the-model) ?event-number ?the-leg
				  ?before-maf-add-additional-info]
		      [add-to-map (sorties ?the-model) ?event-number ?the-sortie
				  ?before-maf-add-additional-info]
		      [add-to-map (movements ?the-model) ?event-number ?the-movement
				  ?before-maf-add-additional-info]
		      [dscs ?the-model mission-builder good]))

(defbehavior-model (maf-add-additional-info compromised)
    :inputs (the-model the-leg the-movement the-sortie event-number)
    :outputs (the-model)
    :prerequisites  ([dscs ?the-leg leg good]
		     [dscs ?the-movement movement good]
		     [dscs ?the-sortie sortie good]
		     [dscs ?the-model mission-builder good])
    :post-conditions ([not [dscs ?the-model mission-builder good]]))

(defsplit maf-more-events? (cmd)
  (build-event (equal ?cmd 'new-event))
  (exit (equal ?cmd 'save-mission)))

(defsplit maf-takeoff? (event)
  (get-additional-info (take-off-event? ?event))
  (exit (not (take-off-event? ?event))))

(define-ensemble maf-save
    :inputs (the-model)
    :outputs ())

(defbehavior-model (maf-save normal)
    :inputs (the-model)
    :outputs ()
    :prerequisites ([dscs ?the-model mission-builder good])
    :post-conditions ([dscs ?the-model mission-builder good]))

(defbehavior-model (maf-save compromised)
    :inputs (the-model)
    :outputs ()
    :prerequisites ([dscs ?the-model mission-builder good])
    :post-conditions ([not [dscs ?the-model mission-builder good]]))
    
;;;;;;;;;;;;;;;;;;;;;;;;;;;;;;;;;;;;;;;;
;;;
;;; attack models
;;; 
;;;;;;;;;;;;;;;;;;;;;;;;;;;;;;;;;;;;;;;;

(define-attack-model maf-attacks
    :attack-types ((hacked-image-file-attack .3) (hacked-code-file-attack .5))
    :vulnerability-mapping ((reads-complex-imagery hacked-image-file-attack)
			    (loads-code hacked-code-file-attack)))

;;; rules mapping conditional probabilities of vulnerability and attacks

(defrule bad-image-file-takeover (:forward)
  if [and [resource ?ensemble ?resource-name ?resource]
	  [resource-type-of ?resource image-file]
	  [resource-might-have-been-attacked ?resource hacked-image-file-attack]]
  then [and [attack-implies-compromised-mode hacked-image-file-attack 
  				?resource hacked .9 ]
	    [attack-implies-compromised-mode hacked-image-file-attack 
	    			?resource normal .1 ]])

(defrule bad-image-file-takeover-2 (:forward)
  if [and [resource ?ensemble ?resource-name ?resource]
	  [resource-type-of ?resource code-memory-image]
	  [resource-might-have-been-attacked ?resource hacked-image-file-attack]]
  then [and [attack-implies-compromised-mode hacked-image-file-attack 
  				?resource hacked .9 ]
	    [attack-implies-compromised-mode hacked-image-file-attack 
	    			?resource normal .1 ]])

(defrule hacked-code-file-takeover (:forward)
  if [and [resource ?ensemble ?resource-name ?resource]
	  [resource-type-of ?resource loadable-files]
	  [resource-might-have-been-attacked ?resource hacked-code-file-attack]]
  then [and [attack-implies-compromised-mode hacked-code-file-attack 
  				?resource hacked .9 ]
	    [attack-implies-compromised-mode hacked-code-file-attack 
	    			?resource normal .1 ]])

(defrule hacked-code-file-takeover-2 (:forward)
  if [and [resource ?ensemble ?resource-name ?resource]
	  [resource-type-of ?resource loadable-files]
	  [resource-might-have-been-attacked ?resource hacked-code-file-attack]]
  then [and [attack-implies-compromised-mode hacked-code-file-attack 
  				?resource hacked .9 ]
	    [attack-implies-compromised-mode hacked-code-file-attack 
	    			?resource normal .1 ]])

;;;;;;;;;;;;;;;;;;;;;;;;;;;;;;;;;;;;;;;
;;;;;
;;;;;      Hacked Code file attacks
;;;;;;;;;;;;;;;;;;;;;;;;;;;;;;;;;;;;;;;


(defrule bad-code-file-takeover (:forward)
  if [and [resource ?ensemble ?resource-name ?resource]
	  [resource-type-of ?resource code-file]
	  [resource-might-have-been-attacked ?resource hacked-code-file-attack]]
  then [and [attack-implies-compromised-mode hacked-code-file-attack 
  											?resource hacked .9 ]
	    [attack-implies-compromised-mode hacked-code-file-attack 
	    										?resource normal .1 ]])

(defrule bad-code-file-takeover-2 (:forward)
  if [and [resource ?ensemble ?resource-name ?resource]
	  [resource-type-of ?resource code-memory-image]
	  [resource-might-have-been-attacked ?resource hacked-code-file-attack]]
  then [and [attack-implies-compromised-mode hacked-code-file-attack 
  						?resource hacked .9 ]
	    [attack-implies-compromised-mode hacked-code-file-attack 
	    					?resource normal .1 ]])
\end{verbatim}

\end{document}